\documentclass[a4paper, 11 pt]{article}
\usepackage[T1]{fontenc}
\usepackage[utf8]{inputenc}
\usepackage[english]{babel}
\usepackage{amsmath}
\usepackage{amssymb}
\usepackage{braket}
\usepackage{mathtools}
\usepackage{dsfont}
\usepackage{subcaption}
\usepackage{booktabs}
\usepackage{yfonts}
\usepackage{setspace}
\usepackage{cite}
\usepackage{verbatim}

\usepackage{amstext} 
\usepackage{array}   
\newcolumntype{C}{>{$}c<{$}} 

\usepackage{geometry}
\usepackage[usenames,dvipsnames]{color}
\usepackage{hyperref}
\hypersetup{
  colorlinks,
  citecolor=Blue,
  linkcolor=Blue,
  urlcolor=Blue}

\numberwithin{equation}{section}

\def\be{\begin{equation}}
\def\ee{\end{equation}}
\def\rme{{\rm e}}
\newcommand{\nn}{\nonumber}
\newcommand{\diff}{\mathrm{d}}
\newcommand{\ii}{\mathrm{i}} 

\def\cc{\mathsf{c}}

\def\kk{{\mathsf{k}}}

\def\nnn{{N}}
\def\D{{\rm D}}
\def \qq{{\tilde q}}
\def \Kt{{\rm K}3}
\def\hatI{{\widehat{I}}}
\def\hatPsi{{\widehat{\Psi}}}
\def\hatZ{{\widehat{Z}}}
\def\korb{{k}}

\newcommand{\FF}{\mathsf{F}}
\newcommand{\cL}{\mathsf{c}}

\newcommand{\AL}{\mathsf{A}}
\newcommand{\AR}{\tilde{\mathsf{A}}}

\begin{document}

\pagestyle{empty}

\begin{center}

$\,$
\vskip 0.5cm

{\Large{\bf Decoupling saddles of the gravitational index and the Farey tail}}

\vskip 1cm

Davide Cassani,${}^{\rm a}$ Alejandro Ruip\'erez,${}^{\rm a, b}$ Enrico Turetta${}^{\rm c}$

\vskip 1cm

\end{center}

\renewcommand{\thefootnote}{\arabic{footnote}}

\begin{center}
{\it ${}^{\rm a}$INFN Sezione di Padova, Via Marzolo 8, 35135, Padova, Italy\\[2mm]
${}^{\rm b}$Dipartimento di Fisica e Astronomia ``Galileo Galilei'', Universit\`a di Padova,\\ Via Marzolo 8, 35135, Padova, Italy\\[2mm]
${}^{\rm c}$Department of Physics and Research Institute of Basic Science, Kyung Hee University, Seoul
02447, Republic of Korea.}

\vskip 3cm

 {\bf Abstract} 
\end{center}

We first construct general saddles of the five-dimensional gravitational supersymmetric index with $S^1\times\mathbb{R}^4$ asymptotics, carrying an arbitrary number of electric charges and Gibbons-Hawking centers. We then uplift the three-charge configurations to type IIB supergravity, obtaining solutions asymptotic to $S^1\times \mathbb{R}^4\times S^1\times {\rm K}3$, carrying D1-D5-P charges. Upon taking a decoupling limit, these solutions yield asymptotically Euclidean AdS$_3\times S^3\times {\rm K}3$ candidate saddles of the dual CFT$_2$ elliptic genus. 
 We focus on the two-center saddles, which involve three integers specifying an orbifold action. Depending on these integers, the saddles may or may not possess a Euclidean horizon, and may be smooth or exhibit orbifold singularities. We give the on-shell action of these two-center saddles in the appropriate ensemble and establish a precise match with the Farey-tail expansion of the elliptic genus at the level of saddle-point actions. The horizonless saddles are the Euclidean version of fractional spectral-flow solutions previously discussed in the literature, corresponding to $({\rm AdS}_3\times S^3)/\mathbb{Z}_k$ orbifolds. The black hole saddles are modular images of the horizonless ones. They generalize the well-known ${\rm SL}(2,\mathbb{Z})$ family of BTZ black holes by placing it in the background of the 
$\mathbb{Z}_k$ orbifold.
We also consider the asymptotically AdS$_3\times S^3$ solutions arising as decoupling limit of five-dimensional black ring and black lens saddles, provide their on-shell action and confirm that they do not contribute to the D1-D5 elliptic genus.

{\noindent }

\newpage
\setcounter{page}{1}
\pagestyle{plain}

\tableofcontents

\newpage 

\section{Introduction}

In recent years, it has been clarified how supersymmetric black holes contribute to supersymmetric indices. The initial developments for   AdS black holes, started in~\cite{Benini:2015eyy,Cabo-Bizet:2018ehj,Choi:2018hmj,Benini:2018ywd}, have also led to new insights into the asymptotically flat case, beginning with~\cite{Iliesiu:2021are}.
The case of AdS$_3$ is somewhat special, having been understood in greater detail for a longer time, largely thanks to the powerful methods available from dual two-dimensional CFTs. Nevertheless, the AdS$_3$ story is consistent with the picture emerged in higher dimensions.
In this paper, we show that certain five-dimensional asymptotically flat saddles admit a natural description within the AdS$_3$/CFT$_2$ framework after taking an appropriate decoupling limit. This provides, on the one hand, a direct connection between these gravitational saddles and their microscopic description in terms of a CFT$_2$ and, on the other hand, a sharper understanding of the relation between the supersymmetric indices computed in AdS$_3$ gravity and in the dual CFT$_2$.

A significant part of the recent progress concerns the gravitational supersymmetric index, namely the gravitational counterpart of the widely studied supersymmetric indices in quantum field theory~\cite{Witten:1982df}. In a supersymmetric theory of gravity, the gravitational index is defined as a partition function computed in the  path integral formulation, with suitable boundary conditions around the Euclidean time circle. These boundary conditions are chosen such that a conserved charge---an angular momentum or an R-charge---acting non-trivially on the supercharge effectively implements the $(-1)^\FF$ insertion ensuring the cancellation between bosonic and fermionic superpartners. Chemical potentials for charges commuting with the supercharge can also be introduced, either through twisted boundary conditions along the Euclidean time circle or, equivalently, by turning on appropriate components of the metric and gauge fields. See~\cite{Cassani:2025sim} for a review.

A major open question regarding the gravitational index, which has attracted considerable attention, is the determination of its semiclassical saddles. These saddles are supersymmetric, non-extremal solutions of the supergravity equations of motion that satisfy the prescribed boundary conditions and have a finite on-shell action. In particular, the Euclidean time circle has finite length, and the corresponding field configurations are generically complex~\cite{Cabo-Bizet:2018ehj}. Throughout the paper, by referring to `Euclidean' saddles we will mean this type of complexified configurations.  

We begin our analysis by considering five-dimensional supergravity with twisted $S^1\times \mathbb{R}^4$ boundary conditions, where $S^1$ denotes the Euclidean time circle and by `twisted' we mean that 
some angular directions of the $S^3$ in $\mathbb{R}^4$
 are fibered through flat connections over $S^1$.
This theory admits a variety of candidate saddles with different topologies, providing a rich setting to explore the gravitational index~\cite{Anupam:2023yns,Hegde:2023jmp,Cassani:2024kjn,Adhikari:2024zif,Boruch:2025qdq,Cassani:2025iix,Boruch:2025sie,Dharanipragada:2026dji} (see also e.g.~\cite{Iliesiu:2021are,H:2023qko,Boruch:2023gfn,Boruch:2025biv} for related studies in four dimensions). For field configurations preserving ${\rm U}(1)^3$ symmetry, the solutions are based on harmonic functions in $\mathbb{R}^3$ with an arbitrary number of aligned centers, and a  classification of their global properties was obtained in minimal supergravity using the rod-structure formalism~\cite{Cassani:2025iix}. In this work, we extend the latter analysis to supergravity coupled to vector multiplets. The corresponding on-shell action, which determines the saddle-point contribution to the path integral, can likewise be calculated using general methods~\cite{Colombo:2025yqy,Cassani:2026teb,BenettiGenolini:2026cdw}.

It is important to emphasize, however, that the existence of a supersymmetric solution satisfying the appropriate boundary conditions does not by itself establish that the solution contributes to the path integral. For this reason, we refer to the configurations discussed above as `candidate' saddles. In the asymptotically AdS case, comparison with the partition function of the dual superconformal field theory (SCFT) can provide valuable information about which saddles contribute. Although in the asymptotically flat case such a dual field-theoretic description is unavailable, in favourable situations one can take a decoupling limit that maps the asymptotically flat solutions to asymptotically AdS configurations, thus restoring the possibility of performing a holographic comparison.

We consider precisely one such situation. We first uplift the five-dimensional candidate saddles carrying three electric charges to type IIB supergravity on $S^1\times \Kt$. This gives configurations that are asymptotic to $S^1  \times \mathbb{R}^4\times S^1 \times \Kt$ and carry D1-D5-P charges. We then take a decoupling limit that yields solutions asymptotic to Euclidean AdS$_3\times S^3\times \Kt$ (where again $S^3$ is twisted over the $S^1\times S^1$ in Euclidean ${\rm AdS}_3$). The $\Kt$ factor does not play any role in the discussion below and we will suppress it henceforth (we also note that it can be replaced by $T^4$ upon introducing a suitably modified index).

In the D1-D5-P setup, the decoupling limit leading to asymptotically ${\rm AdS}_3\times S^3$ solutions is a standard practice \cite{Maldacena:1997re}, for instance it was used in~\cite{Cvetic:1998xh} as a near-horizon limit yielding ${\rm BTZ} \times S^3$ black holes, and it is also commonly applied to horizonless solutions---see e.g.\ the early work~\cite{Maldacena:2000dr}. 
Its application to gravitational index saddles, however, is very recent~\cite{Larsen:2026sav,Nanda:2026mbp,Georgescu:2026uhv} and follows~\cite{Boruch:2025qdq,Boruch:2025sie}, where the same procedure was implemented in a M-theory setup to construct asymptotically ${\rm AdS}_3\times S^2$ index saddles.
While~\cite{Larsen:2026sav,Nanda:2026mbp,Georgescu:2026uhv} discussed the decoupling limit of the specific six-dimensional black string corresponding to the uplift of the five-dimensional BMPV black hole saddle~\cite{Hegde:2023jmp,Anupam:2023yns,Cassani:2024kjn,Adhikari:2024zif,Boruch:2025qdq}, leading to supersymmetric non-extremal ${\rm BTZ}\times S^3$, here we implement it in more general families of saddles, as we are going to illustrate. 

Crucially, when the decoupling limit is applied to gravitational index saddles, the Euclidean time circle remains finite. Consequently, the resulting asymptotically $\mathrm{AdS}_3\times S^3$ configurations are supersymmetric, non-extremal solutions obeying the appropriate boundary conditions for contributing to a supersymmetric index. The on-shell action remains finite and can be directly read from the one of the asymptotically $S^1\times \mathbb{R}^4$ solutions.  We thus obtain natural candidate saddles of the type IIB gravitational index with Euclidean ${\rm AdS}_3\times S^3$ boundary conditions, that in the ensemble of fixed D1-D5 charges can be compared with the elliptic genus of the dual two-dimensional $(4,4)$ SCFT defined on the boundary torus.

The D1-D5 elliptic genus provides a particularly fruitful setting for precision holography. It has several remarkable features: it is known exactly and is invariant under modular transformations of the torus. Moreover, it admits an exact representation in terms of a sum over saddle-point contributions, known as the Farey-tail expansion~\cite{Dijkgraaf:2000fq}; see also~\cite{Moore:2004fg,Kraus:2006wn} for reviews. The large-charge saddles appearing in this expansion are expected to admit a supergravity interpretation.
 An intriguing consequence of the Farey-tail expansion is that the full D1-D5 elliptic genus is determined by the modular images of a finite set of polar contributions. 
The simplest polar states admitting a supergravity description are excitations of ${\rm AdS}_3\times S^3$, with $S^3$ suitably twisted over ${\rm AdS}_3$ so as to realize periodic boundary conditions for both bosonic and fermionic fields around the spatial circle in AdS$_3$. Acting on this saddle with ${\rm SL}(2,\mathbb{Z})$ modular transformations generates a supersymmetric version of the family of BTZ black hole saddles originally given in~\cite{Maldacena:1998bw}, which also contribute to the elliptic genus. These saddles are quotients of Euclidean ${\rm BTZ}\times S^3$, labelled by a pair of coprime integers $(c,d)$, with $c>0$, which, together with integers $(a,b)$ determined by the equation 
$ad-bc =1$,
define an element of ${\rm SL}(2,\mathbb{Z})$. The $c=0$ configuration corresponds to the original polar saddle.
This distinction is also reflected directly in the saddle-point actions: the action of the polar saddle is linear in the chemical potentials, whereas the action of its $(c,d)$ modular images develops a pole at
$c\tau+d=0$. This pole plays a central role in microstate counting, as it is responsible for the exponential growth of states characterizing black hole saddles.

We thus aim to match our general asymptotically Euclidean ${\rm AdS}_3\times S^3$ candidate saddles with large-charge saddles from the Farey-tail expansion of the D1-D5 elliptic genus, through a direct comparison of the saddle-point actions.

We are able to establish a precise match of an infinite family of saddle-point contributions characterized by three integers. 
All these saddles are locally Euclidean ${\rm AdS}_3 \times S^3$ and arise as the decoupling limit of two-center saddles. They differ by global identifications, specified by an orbifold action. Equivalently, a convenient way of distinguishing the geometries is by specifying the circle in the boundary $T^2\times S^3$ that collapses in the bulk and determines how the geometry closes off. 
The three integers are conveniently parameterized as $ck$, $dk$, $p$, where $(c,d)$ are coprime pairs determining elements of ${\rm SL}(2,\mathbb{Z})$ as above, and $k$ is a common factor. The circle  becoming contractible in the bulk can then be specified by providing the vector field $\xi$ that generates it. This reads:
\begin{equation}\label{eq:collapsing_vector_intro}
\xi \,=\, ck\,\partial_{\phi_0}  + dk\, \partial_y  + \left(1-ck -p\right)\,\partial_{\phi_1}+p \,\partial_{\phi_2}\,,
\end{equation}
where $\partial_{\phi_0}$ generates the Euclidean time circle, $\partial_y$ generates the spatial cycle in the torus (that is, the $S^1$ in the $S^1\times \Kt$ compact space), and $\partial_{\phi_1}$, $\partial_{\phi_2}$ generate two circles in $S^3$. These are all Killing vectors in our setup.  
The three integers specify how many times the orbits of $\xi$ wind around the orbits of the three vectors  $\partial_{\phi_0}-\partial_{\phi_1}$, $\partial_{\phi_2}-\partial_{\phi_1}$ and $\partial_{y}$, which in our conventions are the vectors preserving the supercharge used to define the gravitational index. The coefficient of the vector under which the Killing spinor is charged, that in our parameterization is $\partial_{\phi_1}$, is fixed to 1 (up to a sign choice) by regularity in the bulk.

Choosing $c=0,d=1$, we obtain a sub-family of saddles where the Euclidean time circle is not involved in the combination providing the contractible circle, which is then controlled by $k$ and $p$. The corresponding geometry is the one of Euclidean $({\rm AdS}_3\times S^3)/\mathbb{Z}_k$. 
The orbifold acts freely when ${\rm gcd}(p, \korb)= {\rm gcd}(p-1, \korb)=1$, otherwise it has conical singularities at the poles of the $S^3$ fixed by the action of $\xi$, which is then a branched $S^3$.
 These horizonless configurations were previously discussed in a Lorentzian setup in~\cite{Giusto:2012yz} (see also~\cite{Jejjala:2005yu,Berglund:2005vb}), where they were also matched with states in the dual CFT. Here we interpret their Euclidean version as polar state saddles contributing to the elliptic genus. 
 The integer $p$ can be decomposed as $p= \eta k + p'$, where $p'$ gives a finite set of polar states, while $\eta\in\mathbb{Z}$ generates their images under integer spectral flow transformations. In fact, the parameter $\frac{p}{k}=\eta+ \frac{p'}{k}$ can be seen as generating a fractional spectral flow. When $p$ is just a multiple of $k$, the solutions reduce to a class previously studied in~\cite{Lunin:2004uu,Giusto:2004id, Giusto:2004ip, Giusto:2004kj}, where only integer spectral flow transformations were considered.

  The $(c,d)$ modular images of these horizonless saddles give rise to the rest of the saddles that we are able to match with the Farey-tail expansion. For $k=1$ we recover the ${\rm SL}(2,\mathbb{Z})$ family of ${\rm BTZ}\times S^3$ black hole saddles discussed in~\cite{Dijkgraaf:2000fq}. Here, we clarify its asymptotically locally flat origin prior the decoupling limit. The representative $(c=1,d=0)$ of this family is a supersymmetric non-extremal ${\rm BTZ}\times S^3$ black hole, arising as the decoupling limit of the supersymmetric non-extremal BMPV black hole saddle. For $k>1$ we obtain novel ${\rm SL}(2,\mathbb{Z})$ families of black hole saddles, now embedded in $({\rm AdS}_3\times S^3)/\mathbb{Z}_k$. Again, we allow for orbifold singularities at the poles of $S^3$.  

Proving the agreement of the gravitational and CFT  actions for saddles lying above the black hole threshold also requires matching a phase term which is independent of the  chemical potentials. On the field theory side, this phase is generated by the ${\rm SL}(2,\mathbb{Z})$ modular transformation of the polar state, and is a multiple of $a/c$. On the gravity side, it follows from a careful patchwise treatment of the action integral, as discussed in~\cite{Colombo:2025yqy}. This phase term is also necessary to obtain a finite result when taking the limit $c=0$ giving back the polar state action. We show the remarkable agreement between these phases in the field theory and the gravity side. 

We also consider the decoupling limit of saddles with more than two centers, leading to novel  asymptotically ${\rm AdS}_3\times S^3\times \Kt$ solutions to (complexified) type IIB supergravity. In particular, for the three-center saddles describing black ring and black lens saddles, we evaluate the on-shell action in the ensemble of fixed D1-D5 charges. Comparing this with the Farey-tail expansion, we identify an obstruction against interpreting these configurations as saddles of the D1-D5 elliptic genus.

\medskip

The rest of the paper is organized as follows. In section~\ref{sec:multichargesaddles} we illustrate our general construction of asymptotically $S^1\times \mathbb{R}^4$ saddles. In section~\ref{sec:uplift_and_decoupling} we uplift the five-dimensional configurations to type IIB supergravity and discuss the decoupling limit leading to asymptotically ${\rm AdS}_3\times S^3$ solutions, focusing on two-center configurations. In section~\ref{sec:CFT2} we briefly review the elliptic genus of two-dimensional superconformal field theories and the Farey-tail expansion of the D1-D5 elliptic genus. In section~\ref{sec:matchFareyTail}, we match gravitational and CFT saddles by comparing their on-shell actions. In section~\ref{sec:black_ring} we comment on saddles with more than two centers. 
 We conclude in section~\ref{sec: conclusions}. Two appendices contain some technical details of our analysis.


\section{Multi-charge saddles of the 5d gravitational index}
\label{sec:multichargesaddles}
Our primary goal in this section is to revisit and extend a family of saddles of the gravitational index in five dimensions recently constructed in \cite{Cassani:2024kjn,Cassani:2025iix}.\footnote{See also~\cite{Boruch:2025qdq,Boruch:2025sie} for an alternative construction of gravitational index saddles using the 4d/5d uplift.} Our setup here will be $\mathcal{N}=2$ ungauged supergravity coupled to an arbitrary number $n_v$ of vector multiplets. We will construct saddles based on harmonic functions with an arbitrary number of centers, while our previous work \cite{Cassani:2024kjn} only discussed some specific two-center multi-charge saddles, and \cite{Cassani:2025iix} presented a general classification but in the pure supergravity theory ($n_v=0$).  This extension will allow us to describe solutions carrying multiple electric charges and additional discrete data.

We consider the gravitational path integral that computes the following grand-canonical partition function:\footnote{We use a hat to denote grand-canonical partition functions and on-shell actions, so as to distinguish them from the corresponding quantities in a mixed ensemble where some of the electric charges are fixed, to be introduced later on.}
\begin{equation}\label{eq:gravitationalindex}
\hatZ\left(\omega_2, \varphi^I\right)\,=\, {\rm Tr} \, \rme^{2\pi \ii J_1}\rme^{-\beta\{{\cal Q}, \bar{\cal Q}\}+\varphi^I Q_I +\omega_2\left(J_2-J_1\right)}\,.
\end{equation}
Here, ${\cal Q}$ represents a supercharge, $J_1, J_2$ are two  angular momenta rotating orthogonal $\mathbb{R}^2$ planes in $\mathbb{R}^4$, and $Q_I$ are a set of electric charges labeled by the index $I=1,\ldots,n_v+1$. These obey the commutation relations,
\begin{equation}\label{eq:5Dsuperalgebra}
\{{\cal Q, \bar {\cal Q}}\}\,=\, E- {\bar X}^IQ_I\,, \hspace{1cm} [J_i, {\cal Q}]\,=\,\frac{1}{2} {\cal Q}\,, \hspace{5mm} i=1, 2\,, \hspace{1cm} \left[Q_I, {\cal Q}\right]\,=\,0\, ,    
\end{equation}
where $E$ is the energy and ${\bar X}^I$ the asymptotic values of the scalar fields. The parameters $\omega_1, \omega_2$ and $\varphi^I$ correspond to complex chemical potentials conjugated to the angular momenta and the charges. The former are subject to the constraint 
\begin{equation}\label{eq:SUSYconstr}
\omega_1+\omega_2\,=\, 2\pi \ii\, ,
\end{equation}
which precisely gives rise to the $\rme^{2\pi \ii J_1}\,=\, (-1)^{\FF}$ factor in the trace. Since both $Q_I$ and $J_2-J_1$ commute with the supercharge $\cal Q$, standard arguments tell us that the partition function \eqref{eq:gravitationalindex} only receives contributions from states annihilated by both ${\cal Q}$ and $\bar{\cal Q}$. For this reason the partition function \eqref{eq:gravitationalindex} is often called  \emph{the gravitational index}. This is computed by a gravitational path integral with asymptotically locally flat boundary conditions for the fields. More precisely, we impose that the space asymptotes to twisted $S^1\times {\mathbb R^4}$:
\begin{equation}
\diff s^2 \,\longrightarrow\, \diff t_{\rm E}^2+\diff {\tilde r}^2+ {\tilde r}^2 \diff \Omega^2_{(3)}\,, 
\end{equation}
where 
\begin{equation}\label{eq:metricS3}
\diff \Omega_{(3)}^2\,=\, \frac{1}{4}\left[\left(\diff\psi+\cos\theta\diff \phi\right)^2+\diff \theta^2+\sin^2\theta \diff \phi^2\right]\, 
\end{equation}
is the round unitary metric on $S^3$.
The coordinate $\theta$ takes values in the interval $\theta \in [0, \pi]$, and the coordinates $(t_{\rm E}, \psi, \phi)$ satisfy the periodic identifications
\begin{equation}\label{eq:identifications5D}
\left(t_{\rm E}, \psi, \phi\right) \sim \left(t_{\rm E}+\beta, \psi+\ii\omega_-, \phi-\ii\omega_+\right)\sim \left(t_{\rm E}, \psi+2\pi, \phi+2\pi\right)\sim \left(t_{\rm E}, \psi-2\pi, \phi+2\pi\right)\, ,
\end{equation}
where we have introduced the combinations of angular potentials
\begin{equation}
\omega_{\pm}\,=\, \omega_1\pm \omega_2\, ,
\end{equation}
noting that the constraint \eqref{eq:SUSYconstr} amounts to fixing $\omega_+\,=\,2\pi\ii$. The chemical potentials $\varphi^I$ are introduced by fixing the holonomy of the gauge fields around the thermal circle---namely, the circle specified by the first set of identifications in \eqref{eq:identifications5D}. Finally, we must also specify boundary conditions for the fermions. These are taken antiperiodic with respect to each of the U(1)'s in \eqref{eq:identifications5D}. Namely, if ${\cal X}$ represents a generic field, bosonic or fermionic, over which we sum in the path integral, we have that
\begin{equation}\label{eq:bdrycond5D}
 {\cal X}\left(t_{\rm E}, \psi, \phi\right)\sim (-1)^{\FF} {\cal X}\left(t_{\rm E}+\beta, \psi+\ii\omega_-, \phi+2\pi\right)\sim (-1)^{\FF} {\cal X}\left(t_{\rm E}, \psi\pm 2\pi, \phi+2\pi\right)\,,  
\end{equation}
implying
\begin{equation}
{\cal X}\left(t_{\rm E}, \psi, \phi\right)\,\sim\, {\cal X}\left(t_{\rm E}+\beta, \psi+\ii\omega_-+2\pi, \phi\right)\,,  
\end{equation}
consistently with the $(-1)^{\FF}$ in the trace \eqref{eq:gravitationalindex}.

In~\cite{Cassani:2025iix} we provided a systematic classification of saddles of the gravitational index \eqref{eq:gravitationalindex} preserving a U(1)${}^3$ symmetry, working in minimal supergravity. Here, we extend these results to account for the coupling to vector multiplets. Since this is a rather straightforward extension, we shall keep the presentation brief, referring to \cite{Cassani:2025iix} for a detailed study of the resulting geometries.

\subsection{General bubbling saddles}
\label{sec:bubbling_saddles}

The bosonic field content of five-dimensional $\mathcal N=2$ supergravity coupled to $n_v$ vector multiplets consists of the metric $g_{\mu\nu}$, $n_v+1$ vector fields $A^I$ and $n_v+1$ scalars $X^I$, see e.g.~\cite{Bergshoeff:2004kh} for a comprehensive analysis of this theory. The scalars are subject to the cubic constraint
\begin{equation}
C_{IJK}X^I X^J X^K=1\,, 
\end{equation}
where $C_{IJK}$ is a fully-symmetric and constant tensor that determines the couplings of the bosonic action, which  (in Lorentzian signature) is given by
\begin{equation}\label{eq:action_sugra_matter-coupled}
S=\frac{1}{16\pi G_5}\int \left(R\star_5 1-\frac{3}{2}\, a_{IJ}\,\diff X^I \wedge \star_5 \diff X^J-\frac{3}{2}\, a_{IJ} F^I\wedge \star_5 F^{J}
+C_{IJK}A^I\wedge F^{J}\wedge  F^K\right)\, ,
\end{equation}
where
\begin{equation}
\label{eq:defa_{IJ}&X_I}
a_{IJ}=3{X}_I {X}_{J}-2\,{C}_{IJK}X^K\,,  \hspace{1cm} X_I=C_{IJK}X^J X^K\,.
\end{equation}

From now on we restrict to supergravity models for which the scalar manifold is a symmetric space \cite{Gunaydin:1983bi}. This implies the existence of a fully-symmetric and constant tensor $C^{IJK}$ satisfying
\begin{equation}
C^{IJK}C_{J(LM}C_{NP)K}=\frac{1}{27} \, \delta^{I}_{(L}C_{MNP)}\, .
\end{equation}

\paragraph{Local form of the solutions.} We consider a class of supersymmetric solutions in Euclidean signature admitting a ${\rm U}(1)^3$ symmetry. A first ${\rm U}(1)$ is a consequence of supersymmetry and (in the so-called timelike class) is related to the Euclidean time circle introduced before. Assuming a further ${\rm U}(1)$ isometry $\partial_\psi$ commuting with the supercharges, the local form of the metric $g_{\mu\nu}$, vector fields $A^{I}_{\mu}$ and scalars $X^I$ is  \cite{Gauntlett:2002nw, Gauntlett:2004qy}\footnote{The class of solutions we consider are obtained via a Wick rotation $t\to -\ii t_{\rm E}$ of the Lorentzian ones \cite{Gauntlett:2002nw, Gauntlett:2004qy}.} 
\begin{align}
\diff s^2\,&=\,f^2 \left[\diff t_{\rm E} +\ii{\breve\omega}+\ii\omega_{\psi}\left(\diff \psi+\chi\right)\right]^2 + f^{-1}H^{-1}\left(\diff\psi+\chi\right)^2+ f^{-1}H \diff s^2_{\mathbb R^3}\,,\\[1mm]
\label{eq:A^I}
A^{I}\,&=\,-X^{I}f\left(-\ii\,\diff t_{\rm E}+{\breve\omega}\right) + \left(H^{-1}K^{I}-X^I f\omega_{\psi}\right)\left(\diff\psi+\chi\right)+ {\breve A}^{I}-\ii\zeta^{I} \diff t_{\rm E}\, ,\quad \\[1mm]
f^{-1}X_I \,&=\, L_{I} + C_{IJK}\frac{K^JK^K}{H}\,,
\end{align}
where the metric functions $f$ and $\omega_\psi$ are given by
\begin{equation}\label{eq:f}
f^{-3}\,=\,C^{IJK} H_{I} H_J H_K \,, \hspace{1cm} H_I \equiv 3 f^{-1} X_I\,, 
\end{equation}
\begin{equation}
\omega_{\psi}\,=\, M + \frac{3}{2}\frac{L_I K^I}{H}+ C_{IJK}\frac{K^I K^J K^K}{H^2}\,,
\end{equation}
while $\zeta^I$ is a gauge parameter. We note that \eqref{eq:f} follows from the cubic constraint satisfied by the scalars
\begin{equation}
C_{IJK}X^I X^J X^K\,=\, 1 \hspace{5mm}\Leftrightarrow\hspace{5mm} C^{IJK}X_I X_J X_K\,=\, \frac{1}{27}\, .   
\end{equation}
The functions $\left(H, K^I, L_I, M\right)$ and the one-forms $(\chi, {\breve A}^I, \breve\omega)$ live in ${\mathbb R}^3$, and are related via the differential equations 
\begin{equation}\label{eq:eqs_1forms}
\begin{aligned}
\star_{\mathbb R^3}\diff \chi\,=\,& \diff H\,, \hspace{1cm}  \star_{\mathbb R^3}\diff {\breve A}^{I}\,=\,-\diff K^I\,, \\[1mm]
\star_{\mathbb R^3}\diff {\breve \omega}\,=\,& H\diff M-M\diff H+ \frac{3}{2}\left(K^I\diff L_I-L_I\diff K^I\right)\, ,
\end{aligned}
\end{equation}
whose integrability conditions are automatically satisfied if $(H, K^I, L_I, M)$ are harmonic. 

For the solutions we consider here, the harmonic functions $(H, K^I, L_I, M)$ fall within the multi-center class. We now assume a further ${\rm}U(1)$ symmetry, namely we restrict to solutions with U(1)${}^3$ isometry. This  implies that all the centers must be placed along a line in ${\mathbb R}^3$, that we take as the $z$-axis. Given this, we can conveniently describe the solutions using sets of spherical coordinates centered at the origin of ${\mathbb R}^3$ $(r, \theta, \phi)$ and at each of the centers of the harmonic functions $(r_a, \theta_a, \phi)$, as well as the more natural cylindrical coordinates $(\rho, z, \phi)$. These sets of coordinates are related by
\begin{equation}
\rho\,=\, r\sin \theta\,=\,r_a \sin \theta_a\,, \hspace{1cm} z=r\cos\theta\,=\,z_a+r_a\cos \theta_a\,,
\end{equation}
where $z_a$ denotes the position of the $a^{\rm th}$ center. Thus, we have that $a=1, \dots , s$, where $s$ is the total number of centers (sources). This being said, we are ready to present the expressions for the harmonic functions, which are given by \cite{Cassani:2025iix, Cassani:2024kjn}\footnote{The parameters $\kk^I_a$ are  analytic continuations of the parameters $k_a^I$ used in Lorentzian signature in this context, namely $\kk^I_a = \ii k^I_a$. They are such that if $\kk^I_a$ are real, then the metric using the Euclidean time $t_{\rm E}$ is real.}
\begin{equation}
H\,=\, \sum_{a} \frac{h_a}{r_a}\,, \hspace{5mm} K^{I}\,=\,-\ii\sum_{a} \frac{\kk^I_a}{r_a}\,,  \hspace{5mm} L_I\,=\, \ell_{I, 0}+\sum_{a}\frac{C_{IJK}\kk^{J}_{a}\kk^K_a}{h_a r_a}\,, \hspace{5mm}  M\,=\, \frac{\ii}{2}\sum_{a}\frac{C_{IJK}\kk^I_a\kk^J_a \kk^K_a}{h_a^2r_a}\, .
\end{equation}
The parameters are subject to the following constraints,
\begin{equation}\label{eq:constraints}
 \sum_a h_a\,=\, 1\,, \hspace{1cm} \sum_a \kk^I_a\,=\, 0\,,   \hspace{1cm} 27 C^{IJK}\ell_{I, 0}\ell_{J, 0}\ell_{K, 0}\,=\,1\, ,
\end{equation}
so as to ensure asymptotic flatness with standard normalization. The $r$-dependence of the functions $L_I$ and $M$ is fixed by demanding that the solution is nicely capped in the interior. Solving \eqref{eq:eqs_1forms}, we obtain that the one-forms are  locally given by
\begin{equation}\label{eq:oneforms}
\begin{aligned}
\chi\,=\,&\sum_a h_a\cos \theta_a\diff \phi \,, \hspace{1cm}  {\breve A}^I\,=\,\sum_a \,\ii\,\kk^I_a\, \diff\phi\,, \\[1mm]
{\ii\breve \omega}\,=\,& \sum_a \ii w_a \cos\theta_a \diff \phi+\sum_a\sum_{b>a}\frac{C_{ab}}{\delta_{ab}}\left(1+\cos\theta_a\right)\left(1-\frac{r_a+\delta_{ab}}{r_b}\right)\diff \phi \,, 
\end{aligned}
\end{equation}
where 
\begin{equation}
\delta_{ab}\,=\, z_a-z_b\,    
\end{equation}
are the distances between the centers,\footnote{It has been assumed that  $\delta_{ab}>0$ if $b>a$.}
\begin{equation}\label{eq:w_a}
\ii w_a\,=\,-\frac{3}{2}\ell_{I, 0}\kk^I_a-\sum_{b\neq a}\frac{C_{ab}}{|\delta_{ab}|}\,,   
\end{equation}
and 
\begin{equation}\label{eq:defCab}
C_{ab}\,=\,  \frac{h_a h_b}{2} C_{IJK}\left(\frac{\kk^I_a}{h_a}-\frac{\kk^I_b}{h_b}\right)\left(\frac{\kk^J_a}{h_a}-\frac{\kk^J_b}{h_b}\right)\left(\frac{\kk^K_a}{h_a}-\frac{\kk^K_b}{h_b}\right)\,.
\end{equation}

\paragraph{Global regularity.} The solutions just described depend on the parameters $h_a, \kk^I_a, \ell_{I, 0}$ satisfying \eqref{eq:constraints} and the distances between the centers $\delta_{ab}$. 
There are, however, global regularity conditions to be imposed, which we now explain. To this aim, let us first recall that we are demanding the coordinates $(t_{\rm E}, \psi, \phi)$ to be periodically identified as in \eqref{eq:identifications5D}, so that these solutions satisfy the right boundary conditions to contribute to the gravitational index \eqref{eq:gravitationalindex}. The identifications in \eqref{eq:identifications5D} define three circles, with the corresponding ${\rm U}(1)^3$ evolution being generated by the following basis of Killing vectors, 
\begin{equation}\label{eq:basisvectors}
\partial_{\phi_0}\,=\,\frac{\beta}{2\pi}\partial_{t_{\rm E}}+\partial_{\phi}+\frac{\ii\omega_-}{2\pi}\partial_{\psi}\,, \hspace{1cm} \partial_{\phi_1}\,=\, \partial_{\phi}-\partial_{\psi}\,, \hspace{1cm} \partial_{\phi_2}\,=\,\partial_{\phi}+\partial_{\psi}\,,
\end{equation}
which have been normalized so that they have $2\pi$-periodic orbits.\footnote{The adapted $2\pi$-periodic real coordinates $\phi_0,\phi_1,\phi_2$ are defined as $t_{\rm E} = \frac{\beta}{2\pi}\phi_0\,$, $\phi = \phi_0 +\phi_1 + \phi_2\,$, $\psi= \ii\frac{\omega_-}{2\pi} \phi_0 + \phi_2-\phi_1 $. In these coordinates, the asymptotic metric contains fibration terms over the $S^1$ parameterized by $\phi_0$.} The Killing spinor satisfies antiperiodic boundary conditions around each of these circles. The global properties of the solutions can be described using the rod structure formalism, see~\cite{Breunholder:2017ubu} for a Lorentzian analysis using this formalism.
Denoting by $\cal M$ the five-dimensional space, the orbit space ${\cal M}/U(1)^3$ corresponds to a half-plane with boundary the $z$-axis, where the U(1)${}^3$ action degenerates. Indeed, if we denote by ${\cal I}_a=\left[z_a, z_{a+1}\right]$ the rod joining two adjacent centers, we have that the orbits generated by the Killing vector
\begin{equation}\label{eq:rod_vector}
\xi_{{\cal I}_a}\,=\, \partial_{\phi}+2\sum_{b\le a}\ii{w}_{b} \,\partial_{t_{\rm E}}+(2\sum_{b\le a}h_b-1)\,\partial_{\psi}
\end{equation}
contract to zero size at the rod ${\cal I}_a$. Here we are fixing an overall sign choice; this may be relaxed by allowing the coefficient of $\partial_\phi$ to be $\pm 1$. Global regularity demands that $\xi_{{\cal I}_a}$ is a linear combination with integer coefficients of the three basis vectors in \eqref{eq:basisvectors}. A suitable parametrization of these integers is \cite{Cassani:2025iix} 
\begin{equation}\label{eq:rod_vector2}
\xi_{{\cal I}_a}\,=\, n_a \,\partial_{\phi_0}   + \left(1-n_a-p_a\right) \partial_{\phi_{1}}+ p_a \,\partial_{\phi_2}\,.
\end{equation}
Comparing the above expression with \eqref{eq:rod_vector} and using \eqref{eq:basisvectors}, we deduce that the parameters of the solution are related to the boundary data $\beta, \omega_-$ and to the set of integers $n_a,p_a$  by
\begin{equation}\label{eq:regularitycond5D}
\sum_{b\le a}\ii w_b\,=\,\frac{n_a \beta}{4\pi}\,,\hspace{1cm}
\sum_{b\le a}h_b\,=\,\frac{n_a}{2}\left(\frac{\ii\omega_-}{2\pi}+1\right)+p_a\,.
\end{equation}
In turn, an analogous regularity analysis for the gauge fields  yields the following condition 
\begin{equation}\label{eq:regularitycond5D2}
\sum_{b\le a}\kk^I_b\,=\,\frac{n_a\varphi^I}{4\pi}+\frac{q^I_a}{2\ii e^I} \,, 
\end{equation}
where $q^I_a \in \mathbb Z$ are integers characterizing certain flat gauge connections, and $e^I$ is the fundamental charge under $A^I_{\mu}$. Then, we conclude that after imposing boundary conditions and regularity, the solution is entirely determined by the boundary data $\beta, \omega_-, \varphi^I$, together with a set of integers,
\be
n_a, p_a, q^I_a\,\in\,\mathbb{Z}\,, 
\ee
encoding global information about the geometry and the gauge bundle. 

\paragraph{Horizonless solutions.} In the generic case in which at least one of the $n_a$ does not vanish, the combination of U(1) isometries that vanishes at the corresponding rod involves the thermal circle generated by $\partial_{\phi_0}$, thus representing a `Euclidean black hole', or black hole saddle. In contrast, the special class of solutions defined by the condition
\begin{equation}
n_a\,=\,0\, \hspace{1cm} \forall a\, ,
\end{equation}
corresponds to horizonless \emph{thermal solitons}. Namely, when setting $n_a=0$ we recover the Euclidean version of the general class of horizonless solutions studied in \cite{Bena:2005va, Berglund:2005vb, Bena:2007ju} (here with aligned centers), with the Euclidean time being still periodically identified according to \eqref{eq:identifications5D}. To explain how the regularity conditions \eqref{eq:regularitycond5D} and \eqref{eq:regularitycond5D2}  reduce to those found in \cite{Bena:2005va, Berglund:2005vb, Bena:2007ju}, let us note that when all the $n_a$ vanish, the boundary data $\beta, \omega_-$ and $\varphi^I$ drop from \eqref{eq:regularitycond5D} and \eqref{eq:regularitycond5D2}, becoming independent of the parameters of the solution. Therefore, we are left with a one-to-one correspondence between the parameters of the solution and the discrete data $p_a, q^I_a$. More specifically, the first set of equations in \eqref{eq:regularitycond5D} gives rise to the so-called \emph{bubble equations} \cite{Bena:2005va, Bena:2007kg}. This becomes evident after bearing in mind that 
\begin{equation}
\sum_{b\le a}w_b\,=\, 0 \hspace{1cm} \Leftrightarrow \hspace{1cm} w_a\,=\,0 \hspace{5mm} \forall a\,,
\end{equation}
and after using the explicit expression for $w_a$ provided in \eqref{eq:w_a} and \eqref{eq:defCab}, which yields
\begin{equation}
w_a\,=\,0 \, \hspace{5mm} \Leftrightarrow \hspace{5mm} \sum_{b\neq a}\frac{h_a h_b}{|\delta_{ab}|}\left(\frac{\kk^I_a}{h_a}-\frac{\kk^I_b}{h_b}\right)\left(\frac{\kk^J_a}{h_a}-\frac{\kk^J_b}{h_b}\right)\left(\frac{\kk^K_a}{h_a}-\frac{\kk^K_b}{h_b}\right)\,=\,-3\ell_{I, 0}\kk^I_a\,.
\end{equation}
Setting $n_a=0$ in the second of \eqref{eq:regularitycond5D} and in \eqref{eq:regularitycond5D2}, we obtain quantization conditions for the parameters $h_a$ and the fluxes $\kk^I_a$, namely\footnote{The rod structure contains also the non-compact rods ${\cal I}_0\,=\,\left[z_0, z_1\right]$ and ${\cal I}_s\,=\,\left[z_s, z_{s+1}\right]$ joining the first and last centers with $z_0\,=\,+\infty$ and $z_{s+1}\,=\,-\infty$. The Killing vectors degenerating there are $\partial_{\phi_1}$ and $\partial_{\phi_2}$, respectively. This implies $n_0\,=\,n_s\,=\,0$, $p_s\,=\,1-p_0\,=\,1$.}
\begin{equation}
h_a\,=\, p_a -\sum_{b<a}p_b\,, \hspace{1.3cm} 2\ii e^I\kk^I_a\,=\, q^I_a-\sum_{b<a}q^I_b\,, \hspace{1.3cm} a=1, \dots, s\,. 
\end{equation}

\paragraph{Asymptotic charges.} The expressions for the charges and angular momenta in terms of the parameters of the solution are:
\begin{equation}\label{eq:5Dcharges}
Q_I\,=\,\frac{3\pi}{G_5}\sum_a \frac{C_{IJK}\kk^J_a \kk^K_a}{h_a}\,, \hspace{5mm} J_+\,=\,-\frac{3\pi \ii}{G_5}\ell_{I, 0}\sum_a\kk^I_az_a\,, \hspace{5mm} J_-\,=\,-\frac{\pi \ii}{G_5}C_{IJK}\sum_a \frac{\kk^I_a\kk^J_a \kk^K_a}{h^2_a}\,,
\end{equation}
where
\begin{equation}
J_{\pm}\,=\, \frac{J_1\pm J_2}{2}\,.
\end{equation}
Supersymmetry then fixes the mass $E$ in terms of the charges as 
\begin{equation}
E\,=\,{\bar X}^IQ_I\,=\,  27 C^{IJK}\ell_{J, 0} \ell_{K, 0} Q_I\,. 
\end{equation}


\subsection{Details on two-center saddles: action, entropy and discrete data}
\label{sec:2centresaddle}

The on-shell action of the saddles we have just described has been computed in a series of works \cite{Cassani:2024kjn, Colombo:2025ihp, Cassani:2025iix, BenettiGenolini:2025icr, Colombo:2025yqy}. Let us discuss it, focusing on solutions with two centers, which will be our focus for most of the paper.

Making use of the notation introduced in \cite{Cassani:2024kjn, Cassani:2025iix}, we refer to the two centers by north ($N$) and south ($S$) poles. Since there is just one compact rod joining them, from now on we drop the label from the integers associated to it, simply denoting them by $n, p, q^I$. From \eqref{eq:regularitycond5D} and \eqref{eq:regularitycond5D2}, one obtains that the parameters of the solution are related to these integers and the boundary data by 
\begin{equation}\label{eq:expr_parameters_1}
h_N \,=\, n \frac{\omega_2}{2\pi\ii}+p\,,\qquad h_S \,=\, n\frac{\omega_1}{2\pi\ii} + 1-n-p\,,\qquad \kk^I_N \,=\, -\kk^I_S \,=\, \frac{\ii}{2}\left(n\frac{\varphi^I}{2\pi\ii}-\tilde q^I\right)\,,
\end{equation}
and 
\begin{equation}\label{eq:expr_parameters_2}
\delta^{-1} \,=\, -\frac{h_N^2h_S^2}{2\pi}\frac{6\pi\ell_{I,0}\kk_N^I + n\beta}{C_{IJK}\kk^I_N\kk^J_N \kk^K_N}\,,
\end{equation}
where we have defined 
\begin{equation}
\qq^I\,=\,\frac{q^I}{e^I}\,.
\end{equation}

The Killing vector \eqref{eq:rod_vector2} degenerating at the compact rod is given by 
\begin{equation}\label{eq:xi_N}
\xi \,=\, n\,\partial_{\phi_0}  + \left(1-n-p\right)\,\partial_{\phi_1}+p \,\partial_{\phi_2}\,.
\end{equation}
As already explained, we can interpret the $n\neq 0$ case as describing a family of Euclidean black holes labeled by the integers $n, p, q^I$. The Euclidean horizon corresponds to the fixed locus of the Killing vector \eqref{eq:xi_N}.
Among this family of black hole saddles, there is one corresponding to the supersymmetric non-extremal version of the BMPV black hole~\cite{Breckenridge:1996is}, given by the choice $n=1$, $p\,=\,q^I\,=\,0$  \cite{Cassani:2024kjn}. 
Instead, the $n= 0$ case gives the thermal version of the horizonless solutions discussed in \cite{Giusto:2004id, Giusto:2004ip, Giusto:2004kj, Giusto:2012yz} (see also~\cite{Jejjala:2005yu,Berglund:2005vb}). The description in terms of harmonic functions can be found e.g.\ in~\cite{Giusto:2012yz}. 

The fixed locus of the Killing vector \eqref{eq:xi_N} is an orbifold of $S^3$.  When ${\rm gcd}(n,p)={\rm gcd}(n,p-1)=1$, the orbifold is freely acting and gives the smooth lens space 
$S^3/\mathbb Z_{|n|}$
\cite{Cassani:2025iix}.
When either one (or both) of ${\rm gcd}(n,p)$ or ${\rm gcd}(n,p-1)$ is different from 1, the five-dimensional geometry has a $\mathbb{Z}_{{\rm gcd}(n,p)}$ or a  $\mathbb{Z}_{{\rm gcd}(n,p-1)}$ orbifold singularity at either one (or both) of the poles of $S^3/\mathbb Z_{|n|}$, which is thus a branched lens space. This also applies to the horizonless solutions obtained by taking $n=0$: in this case ${\rm gcd}(0,p)=|p|$, ${\rm gcd}(0,p-1)=|p-1|$, and the five-dimensional geometry has  a $\mathbb{Z}_{|p|}$ orbifold singularity at a pole and a  $\mathbb{Z}_{|p-1|}$ singularity at the other pole~\cite{Giusto:2004kj}. We anticipate that these singularities are partially resolved when uplifting the solutions to type IIB supergravity: the uplift involves an $S^1$ that is non-trivially fibred over the five-dimensional geometry, hence we will need to come back to the orbifold structure again after uplifting the solutions.

\paragraph{On-shell action.} The saddle-point contribution of two-center saddles to the partition function \eqref{eq:gravitationalindex} is given by their on-shell action \cite{Cassani:2025iix,Colombo:2025yqy}:
\begin{equation}
\label{eq:2centregeneral}
    \hatI \,=\, \frac{\pi}{4G_5}\,\frac{C_{IJK}\left(n\varphi^I - 2\pi \ii \tilde q^I\right)\left(n\varphi^J - 2\pi \ii \tilde q^J\right) \left( n \varphi^K - 2\pi \ii \tilde q^K\right)}{n\left(n\omega_1 + 2\pi \ii \left(1-n-p\right)\right)\left(n\omega_2 + 2\pi \ii p\right)} + 2\pi \ii\, \hatPsi\,,
\end{equation}
where
\be\label{eq:phaseterm}
\hatPsi \,=\, \frac{\pi}{4G_5}\,w_0^2(w_0^1)^2(n+p-1) \,\frac{C_{IJK}\tilde q^I \tilde q^J \tilde q^K}{n}\,.
\ee
The first term was derived in \cite{Cassani:2025iix} in minimal five-dimensional supergravity, and extended in \cite{Colombo:2025yqy} to the matter-coupled case (it also corresponds to the ungauged limit of expressions first derived for AdS$_5$ black holes~\cite{Cabo-Bizet:2018ehj,Aharony:2021zkr}). The purely imaginary term $\hatPsi$, which is independent of the chemical potentials, was added in \cite{Colombo:2025yqy} as the result of a careful patchwise treatment of the action integral, different gauge patches being needed when the flat connections $\tilde q^I$ are turned on. See \cite{Colombo:2025yqy} and in particular section\ 5.1.1 therein for details.\footnote{The precise map with~\cite[eq.~(5.26)]{Colombo:2025yqy} is:
$
\ (M^I)_{\rm there}=- \frac{\tilde q^I}{n}\,,\ \ (\kappa)_{\rm there}=1\,,\ \ (x)_{\rm there} = 1-n-p\,,\ \ (y)_{\rm there} = p\,
$.}
The quantities $w_0^1,w_0^2$ are components of vectors $w^a$ in the basis~\eqref{eq:basisvectors}, defined as $w^a = w^a_0\partial_{\phi_0}+ w^a_1\partial_{\phi_1}+ w^a_2 \partial_{\phi_2}$, $a=1,2$. They satisfy the equations
\begin{equation}
\label{def:w}
   1 = w_1^1 n - w_0^1  \left(1-n-p\right)\,,\qquad\quad 1 = w_2^2 n - w_0^2 p\,.
\end{equation}
If ${\rm gcd}(n,p)= {\rm gcd}(n,p-1)=1$, then $w_0^1, w_0^2, w_1^1, w_2^2$ are taken integers, their existence being guaranteed by Bezout's lemma. Otherwise, in general, one should take
\be\label{eq:quantization_w}
w_0^1,w_1^1 \,\in\, \frac{\mathbb{Z}}{{\rm gcd}(n,p-1)}\,, \qquad\quad  w_0^2,w_2^2 \,\in\, \frac{\mathbb{Z}}{{\rm gcd}(n,p)}\,.
\ee
Being independent of the chemical potentials, the term $\hatPsi$  does not affect the conjugate charges of the solution. It just contributes as a phase to $\rme^{-\hatI}$. Later on, in section~\ref{sec:matchFareyTail}, we will see that this phase is in fact needed in order to precisely match the action predicted by modularity of the AdS$_3$/CFT$_2$ partition function.
 
\paragraph{Entropy.} While the on-shell action $\hatI$ can be seen as a thermodynamic function in a grand-canonical ensemble where the chemical potentials are the independent variables and the constraint $\omega_1+\omega_2\,=\,2\pi\ii$ is imposed, we can also consider an ensemble where  all the charges apart for $J_1$ are the independent variables. Note that we cannot fix $J_1$ because we are fixing the conjugate chemical potential to realize the $(-1)^\FF$ insertion.
The corresponding thermodynamic function, obtained by taking a  Legendre transform of $\hatI$ with respect to the available chemical potentials $\omega_2$ and $\varphi^I$,
is defined as
\begin{equation}\label{eq:qsr}
{\cal S}(J_-,Q_I)\,=\, {\rm ext}_{(\omega_2,\varphi^I)} \left[ -\hatI(\omega_2,\varphi^I) - \omega_1 J_1 - \omega_2 J_2 - \varphi^I Q_I\right]\,, \qquad \text{with}\ \ \omega_1 \,=\, 2\pi\ii-\omega_2\,,\quad
\end{equation}
and gives the `indicial entropy' 
\begin{equation}
{\cal S} + 2\pi\ii J_1 \,=\, {\rm ext}_{(\omega_2,\varphi^I)} \left[ -\hatI\,(\omega_2,\varphi^I)  + 2\omega_2 J_- - \varphi^I Q_I\right]\,,\end{equation}
    where $J_1$ should be seen as a saddle-point (average) value. 
   In this way, we obtain a saddle-point contribution to a partition function computing indicial degeneracies, given by an index-trace over microstates that have fixed $J_-$, $Q_I$  but can carry any $J_1$, 
   \be\label{eq:quasimicroZ}
   {\rm Tr}_{(J_-,\,Q_I \, \text{fixed})}\, \rme^{2\pi\ii J_1}\,=\, d_{\rm bosonic}(J_-,Q_I) - d_{\rm fermionic}(J_-,Q_I)\,.
   \ee

 Implementing the Legendre transform of the on-shell action~\eqref{eq:2centregeneral} for $n\neq0$, we obtain 
\begin{equation}
\label{eq:entropy2centergeneral}
\begin{aligned}
    {\cal S} +2\pi\ii J_1 &\,=\, \frac{4\sqrt{\pi\, G_5}}{n}\sqrt{C^{IJK}Q_I Q_J Q_K - \frac{\pi}{4G_5}J_-^2} \, -\, \frac{2\pi \ii}{n} \left(\left(2p-1\right) J_- + \tilde q^I Q_I + n\hatPsi\right) ,
    \end{aligned}
\end{equation}
where $\hatPsi$ is the same as in \eqref{eq:phaseterm}.
 If we now set $n = 1$, $p=q^I =0$, we should recover the entropy of the BMPV black  hole. Indeed, one gets
\begin{equation}
{\cal S}+2\pi\ii J_+\,=\,  4\sqrt{\pi G_5}\sqrt{C^{IJK}Q_I Q_J Q_K - \frac{\pi}{4G_5}J_-^2}\,,
\end{equation}
which agrees with the BMPV extremal entropy~\cite{Breckenridge:1996is} after the condition $J_+=0$ is imposed. As explained in \cite{Anupam:2023yns, Cassani:2024tvk}, the vanishing of $J_+$ is equivalent to the extremality condition, $\beta \to \infty$, for this saddle.

\paragraph{Horizonless saddles.} 
We can derive the on-shell action of two-center horizonless saddles
by taking a formal $n\to 0$ limit of the action~\eqref{eq:2centregeneral}.
Although the first and main term in the action is divergent in this limit, remarkably the phase term $\hatPsi$ precisely compensates this divergence, so that the limit is in fact finite (this can be seen by noting that for $n=0$, $w_0^1 = 1/(p-1)$ and $w_0^2 = -1/p$ in \eqref{eq:phaseterm}). We thus have a further reason for including the constant phase $\hatPsi$. The $n\to 0$ limit then gives
\begin{equation}
\label{eq:horizonlessgeneral}
\hatI_{n\to 0} \,=\, \frac{\pi}{4G_5}\frac{C_{IJK}\,\qq^I \qq^J}{p(1-p)}\left[3\varphi^K + \qq^K \left(\frac{\omega_1-2\pi\ii}{1-p}+\frac{\omega_2}{p}\right)\right]\,,
\end{equation}
which we will take as the contribution of the two-center solitons to the gravitational index.

Let us make a few comments about the formula above. First, we emphasize again that for horizonless solutions the chemical potentials $\omega_1,\omega_2,\varphi^I$ are  unrelated to the solution parameters, since the Euclidean time circle does not collapse anywhere. The action can therefore be determined up to an integration constant by demanding that its variation with respect to each chemical potential gives (minus) the conjugate charge, see also the discussion in section 5.3 of~\cite{Colombo:2025yqy}. The action \eqref{eq:horizonlessgeneral} does have this property. However, since in this paper we are focusing on the gravitational index partition function, we need each candidate saddle to satisfy the corresponding boundary conditions, which amounts to imposing the constraint $\omega_1+\omega_2 = 2\pi \ii$. We do so in \eqref{eq:horizonlessgeneral} a posteriori.

We also note that the integration constant, namely the additional  term independent of the chemical potentials, would be interpreted as minus the entropy of the solution, so it should be at most a trivial phase for horizonless solutions. In \eqref{eq:horizonlessgeneral}, this constant is determined by the $n\to 0$ procedure above, and it is such that
\begin{equation}\label{eq:S2csol}
{\cal S}+2\pi\ii \, J_1\,=\,0\,, \hspace{1cm} J_1\,=\, -\frac{\pi}{4 G_5}\frac{C_{IJK}\qq^I\qq^J\qq^K}{p\left(1-p\right)^2}\,
\end{equation}
(where the expression for $J_1$ follows from~\eqref{eq:5Dcharges}, using~\eqref{eq:expr_parameters_1} and~\eqref{eq:expr_parameters_2} with $n=0$).
Thus, we see that the entropy is a trivial phase when $J_1\in \mathbb Z$.  Later, we will see how this condition arises when embedding these solutions in type IIB string theory. As it turns out, upon taking a decoupling limit, the thermal two-center solitons correspond to the gravitational dual of (a subset of) the polar states contributing to the elliptic genus of the dual D1-D5 CFT${}_2$. Since the latter are individual CFT states, the soliton must carry integer quantized angular momenta $J_1, J_2$. We conclude that this solution contributes like 1 (at the classical level) to the trace \eqref{eq:quasimicroZ}.

Finally, we would like to conclude this section commenting on an observation made in~\cite{Cassani:2024kjn, Cassani:2025iix}, where it was argued that the Lorentzian solitons can be obtained from the BMPV saddle ($n=1$ and $p=q^I=0$) by sending $\beta \to 0$ and performing suitable analytic continuations. While such limit does not affect the on-shell action, which is independent of $\beta$, it affects the global identifications \eqref{eq:identifications5D}, requiring the chemical potentials to be quantized as multiples of $2\pi \ii$. The limit $n\to 0$ considered here locally gives the same configuration as the $\beta\to 0$ limit, but it leaves the global identifications \eqref{eq:identifications5D} unchanged, namely the $n\to 0$ solutions exist for generic values of the chemical potentials. Fixing the chemical potentials in \eqref{eq:horizonlessgeneral} in the same way as in the $\beta\to 0$ limit, so as to be able to compare with \cite{Cassani:2024kjn, Cassani:2025iix}, one finds that the on-shell actions in the $n\to0$ and $\beta\to0$ limits are related by $\hatI_{n\to 0}\,=\, \hatI_{\beta \to 0}-2\pi\ii J_1$. Thus, we find that the respective saddle-point contributions to the path integral agree since $J_1\in \mathbb Z$. 


\section{Uplift to type IIB on $\boldsymbol{S^1\times M_4}$ and decoupling limit}
\label{sec:uplift_and_decoupling}

A limitation of the asymptotically flat gravitational index is the lack of an obvious microscopic realization to compare with, in particular the lack of a holographically dual field theory. A strategy to go around this is to take a decoupling limit such that the asymptotically flat region is excised from the geometry, leading to asymptotically AdS solutions \cite{Maldacena:1997re}. This was applied both to black hole and horizonless solutions long ago---see  \cite{Cvetic:1998xh} and e.g.  \cite{Maldacena:2000dr}, respectively---while the application to supersymmetric non-extremal configurations has begun recently with~\cite{Boruch:2025qdq,Boruch:2025sie}.  In this section, we discuss the decoupling limit for the solutions studied in section~\ref{sec:multichargesaddles}. To this aim, we consider a specific supergravity model within the general class studied in the previous section. This is the STU model, which arises as a consistent truncation of type IIB supergravity on $S^1 \times M_4$, where $M_4$ can be either $T^4$ or ${\rm K}3$, and describes a sector of the dynamics of the D1-D5 brane system in this geometry. Uplifting our solutions to ten dimensions will allow us to take a decoupling limit leading to asymptotically AdS${}_3\times S^3\times M_4$ solutions. This will set the stage for a comparison with the dual CFT${}_2$, which will be discussed in section~\ref{sec:matchFareyTail}. 

For this reason, in this section we will perform a change of thermodynamical ensemble with respect to the previous section. Namely, we will work in a mixed ensemble where we keep the D5- and D1-brane charges fixed. These will correspond to $Q_1$ and $Q_2$ in our setup.

\subsection{Type IIB uplift}\label{sec:STU}

The STU model has $n_v=2$ vector multiplets, and it is characterized by the choice $C_{IJK}\,=\, \frac{1}{6}|\epsilon_{IJK}|$. Being a consistent truncation, any solution of this model can be uplifted to a solution of type IIB supergravity. The embedding can be found e.g.~in \cite{Elvang:2004ds}, and it is given by
\begin{equation}\label{eq:upliftformulae}
\begin{aligned}
\diff s^2_{\rm IIB}\,&=\,\left(X^3\right)^{1/2}\diff s^2 + \left(X^3\right)^{-3/2}\left(\diff y -A^3\right)^2+ X^1 \left(X^3\right)^{1/2}\diff s^2_{M_4}\,, \\[1mm]
 F_{(3)}\,&=\, g_s^{-1}\left[\left(X^1\right)^{-2}\star_5 F^1 +F^2\wedge \left(\diff y-A^3\right)\right]\,,\\[1mm]
\rme^{2\phi}\,&=\, g_s^2\frac{H_2}{H_1}\, ,
\end{aligned}
\end{equation}
where $y\sim y +2\pi R_y$ is the coordinate parametrizing $S^1$, and we are using the Hodge star in Lorentzian signature. The internal space $M_4$ can be either $T^4$, or ${\rm K}3$. 
We further note that we are fixing the asymptotic values of the scalars as
\begin{equation}
\lim_{r\to \infty} X^I\,=\, 1 \hspace{1cm} \Leftrightarrow \hspace{1cm} \ell_{I, 0}\,=\, \frac{1}{3}\, ,    
\end{equation}
which does not affect the on-shell action. Explicitly, we can write the type IIB metric (in the string frame) as
\begin{equation}\label{eq:typeIIBmetric}
\begin{aligned}
\diff s^2_{\rm IIB}\,=\,& \left(H_1 H_2\right)^{-1/2}\left[H_3^{-1}\left(\diff t_{\rm E} + \ii\omega\right)^2 + H_3\left(\diff y-A^3\right)^2\right]+ \left(H_1 H_2\right)^{1/2} \left[H^{-1}\left(\diff \psi+\chi\right)^2\right.\\[1mm]
&\left.+H\diff s^2_{\mathbb R^3}\right]+ \left(\frac{H_2}{H_1}\right)^{1/2} \, \diff s^2_{M_4}\,,
\end{aligned}
\end{equation}
where 
\begin{equation}
H_1\,=\, 3 L_1+ \frac{K^2 K^3}{H}\,, \hspace{1cm} H_2\,=\, 3 L_2+ \frac{K^1 K^3}{H}\,, \hspace{1cm}
H_3\,=\, 3 L_3+ \frac{K^1 K^2}{H}\,.
\end{equation}
If we fix the gauge parameter in \eqref{eq:A^I} to  $\zeta^3\,=\,1$ for the KK vector, so that it vanishes asymptotically, then we have that the periodic identifications of the coordinates are given by
\begin{equation}\label{eq:periodicidentifications}
\begin{aligned}
\left(t_{\rm E}, y, \psi, \phi\right)\,\sim&\, \left(t_{\rm E} +\beta, y-\ii \varphi^3-\ii\beta,\psi+\ii\omega_-, \phi+2\pi\right)\sim \left(t_{\rm E}, y+2\pi R_y, \psi, \phi\right) \\[1mm]
\,\sim& \,\left(t_{\rm E}, y, \psi +4\pi, \phi\right)\sim \left(t_{\rm E}, y, \psi+2\pi, \phi+2\pi\right)\,,
\end{aligned}
\end{equation}
where we note that the shift of  $y$ when going once around the thermal circle follows from the geometrization of the U(1) gauge transformations of the KK vector, $A^3$. Indeed, in appendix~\ref{app:10Dregularity} we show that a regularity analysis in ten dimensions yields the same conditions \eqref{eq:regularitycond5D} and \eqref{eq:regularitycond5D2} we derived in five dimensions, after identifying the fundamental charges under each of the vectors as
\begin{equation}\label{eq:fundamental_charges}
e^1\,=\,\frac{v R_y}{g_s \alpha'} \,, \hspace{1cm} e^2\,=\,\frac{R_y}{g_s \alpha'}\, , \hspace{1cm} e^3\,=\,\frac{1}{R_y}\,,
\end{equation}
where we have defined the dimensionless $M_4$ volume
\begin{equation}
v\,\equiv\,\frac{{\rm{Vol}}_{M_4}}{\left(2\pi\right)^4\alpha'{}^2}\,.
\end{equation}
The identifications \eqref{eq:fundamental_charges} stand from the fact that the charges $Q_1, Q_2$ and $Q_3$ are associated to D5 branes, D1 branes and KK momentum $\rm P$, respectively. Given this, it is natural to introduce the integer-quantized charges $N_{\rm D1}$, $N_{\rm D5}$, $N_{\rm P}$ via\footnote{While for $M_4=T^4$ $N_{\rm D1}$ and $N_{\rm D5}$ denote the actual numbers of D1- and D5-branes, for $M_4 = {\rm K}3$ by $N_{\rm D1}$ we denote the {\it net} number of D1-branes, including a negative shift due to the D1-charge induced on the worldvolume of the D5-branes by curvature coupling. Specifically, if $n_{\rm D1}$ and $n_{\rm D5}$ denote the actual number of D1- and D5-branes for $M_4={\rm K}3$, then $N_{\rm D1}= n_{\rm D1}-n_{\rm D5}$ and $N_{\rm D5}= n_{\rm D5}$.}
\begin{equation}\label{eq:integer_charges}
Q_1\,=\,\frac{v R_y N_{\D5}}{g_s \alpha'} \,, \hspace{1cm} Q_2\,=\,\frac{R_y N_{\D1}}{g_s \alpha'}\,, \hspace{1cm} Q_3\,=\,\frac{N_{\rm P}}{R_y}\,.
\end{equation}
Finally, we will need the relations between Newton constants in five, six and ten dimensions,
\begin{equation}\label{eq:Newtons_const}
G_5\,=\,\frac{G_6}{2\pi R_y}\,=\,\frac{G_{10}}{\left(2\pi\right)^5 v \,\alpha'{}^2 R_y}\,=\, \frac{\pi g_s^2\alpha'{}^2}{4 v R_y}\,.   
\end{equation}

\subsection{Decoupling limit}
\label{sec:decoupling}

The decoupling limit we are after amounts to rescaling the coordinates and parameters of the solution as
\begin{equation}\label{eq:decouplinglimit}
t_{\rm E} \to \, \frac{t_{\rm E}}{\epsilon}\,, \hspace{4mm} y \to \,\frac{y}{\epsilon}\,, \hspace{4mm} {\vec x}\to \epsilon^2{\vec x}\,, \hspace{4mm}\kk^{1, 2}_a \to  \epsilon\, \kk^{1, 2}_a\,, \hspace{4mm} \kk^3_a \to \frac{\kk^3_a}{\epsilon}\,, \hspace{4mm} \beta \to \frac{\beta}{\epsilon}\,, \hspace{4mm}R_y\to \frac{R_y}{\epsilon} \,,
\end{equation}
and then send $\epsilon \to 0$. From now on, we fix
\begin{equation}\label{eq:Ry=1}
    R_y\,=\,1\,,
\end{equation}
so that the new $y$ coordinate is $2\pi$-periodic.
The net effect of this decoupling limit is to remove the constant term in the harmonic functions $L_{1}$ and $L_{2}$. Indeed, in terms of the rescaled parameters and coordinates, the harmonic functions characterizing the decoupled solutions are given by
\begin{equation}
\begin{aligned}
H\,=\,& \sum_{a} \frac{h_a}{r_a}\,, \hspace{1cm} {K}^{I}\,=\,-\ii \sum_{a} \frac{\kk^I_a}{r_a}\,,   \hspace{5mm}(I=1, 2,3)\hspace{1cm} M\,=\, \frac{\ii}{2}\sum_{a}\frac{ \kk^1_a\kk^2_a \kk^3_a}{h_a^2r_a}\, , \\[1mm]
L_1\,=\,& \sum_{a}\frac{\kk^{2}_{a}\kk^3_a}{3h_a r_a}\,,\hspace{1cm}L_2\,=\, \sum_{a}\frac{\kk^{1}_{a}\kk^3_a}{3h_a r_a}\,,\hspace{1cm}L_3\,=\, \frac{1}{3}+\sum_{a}\frac{\kk^{1}_{a}\kk^2_a}{3h_a r_a}\,,
\end{aligned}
\end{equation}
and the local expressions for the one-forms are the same as in \eqref{eq:oneforms}, with the only difference that the coefficients $w_a$ are now given by 
\begin{equation}
\ii w_a\,=\,-\frac{\kk^3_a}{2}-\sum_{b\neq a}\frac{C_{ab}}{|\delta_{ab}|}\,,   
\end{equation}
where we have again that
\begin{equation}
C_{ab}\,=\,  \frac{h_a h_b}{2} \left(\frac{\kk^1_a}{h_a}-\frac{\kk^1_b}{h_b}\right)\left(\frac{\kk^2_a}{h_a}-\frac{\kk^2_b}{h_b}\right)\left(\frac{\kk^3_a}{h_a}-\frac{\kk^3_b}{h_b}\right)\,.
\end{equation}
Furthermore, the new coordinates are still periodically identified as in \eqref{eq:periodicidentifications}, noting that we imposed \eqref{eq:Ry=1}.  The relation between the parameters of the solution and the boundary conditions can be retrieved from \eqref{eq:regularitycond5D} and \eqref{eq:regularitycond5D2}, after using \eqref{eq:fundamental_charges}. This yields
\begin{equation}\label{eq:regularitycondIIB}
\sum_{b\le a}\ii w_b\,=\,\frac{n_a \beta}{4\pi}\,,\hspace{1cm}
\sum_{b\le a}h_b\,=\,\frac{n_a}{2}\left(\frac{\ii\omega_-}{2\pi}+1\right)+p_a\,,
\end{equation}
and
\begin{equation}\label{eq:regularitycondIIB2}
\begin{aligned}
\sum_{b\le a}\kk^1_b\,=\,&\frac{n_a\varphi^1}{4\pi}+\frac{g_s \alpha'q^1_a}{2\ii v} \, , \hspace{1cm}
\sum_{b\le a}\kk^2_b\,=\,&\frac{n_a\varphi^2}{4\pi}+\frac{g_s \alpha'q^2_a}{2\ii} \, , \hspace{1cm}
\sum_{b\le a}\kk^3_b\,=\,&\frac{n_a\varphi^3}{4\pi}+\frac{q^3_a}{2\ii} \, .
\end{aligned}
\end{equation}
When at least one of the $n_a$ is different from zero, we should understand the first two equations above as fixing (in particular) the chemical potentials $\varphi^1$ and $\varphi^2$ in terms of the charges $Q_1$ and $Q_2$ together with the remaining thermodynamical variables and discrete data. We will further discuss this point for the two-center saddles in section~\ref{sec:decoupling2center}. In the meantime, we provide the expressions for the charges and angular momenta in terms of the parameters of the solution:
\begin{equation}\label{eq:charges_STU}
\begin{aligned}
Q_{1}\,=\,& \frac{2\pi^2}{G_6} \sum_a \frac{\kk^2_a \kk^3_a}{h_a}\,, \hspace{1cm} Q_{2}\,=\, \frac{2\pi^2}{G_6} \sum_a \frac{\kk^1_a \kk^3_a}{h_a}\,, \hspace{1cm} Q_{3}\,=\,\frac{2\pi^2}{G_6} \sum_{a}\frac{\kk^1_a \kk^2_a}{h_a}\,, \\[1mm]
J_{+}\,=\,& -\frac{2\ii\pi^2}{G_6} \sum_a \kk^3_a z_a\,, \hspace{1cm} J_{-}\,=\, -\frac{2\ii\pi^2}{G_6} \sum_a\frac{\kk^1_a\kk^2_a \kk^3_a}{h_a^2}\, . 
\end{aligned}
\end{equation}

\subsection{Asymptotic behavior and boundary conditions after the decoupling limit}
\label{sec:AdS3xS3}

 In order to show that the decoupling limit \eqref{eq:decouplinglimit} leads to asymptotically AdS${}_3\times S^3$ solutions, we spell out the leading asymptotic behavior of the functions and one-forms appearing in the metric. We have that  
\begin{equation}
 H_{1} \underset{r\to \infty}{\longrightarrow} \,\frac{G_6Q_{1}}{2\pi^2 r}\,, \hspace{1cm}  H_{2} \underset{r\to \infty}{\longrightarrow} \frac{G_6 Q_2}{2\pi^2 r}\,, \hspace{1cm} H_3\underset{r\to \infty}{\longrightarrow} 1\,, \hspace{1cm} H\,\underset{r\to \infty}{\longrightarrow}\, \frac{1}{r} \,, 
\end{equation}
and
\begin{equation}
\begin{aligned}
 \chi \underset{r\to \infty}{\longrightarrow}\,& \cos \theta\, \diff \phi \,, \\[1mm]
 \omega \underset{r\to \infty}{\longrightarrow}\,&-\frac{G_6 J_-}{4\pi^2 r}\left(\diff \psi+\cos\theta \diff \phi\right)+\frac{G_6 J_+}{4\pi^2 r}\left(\diff\phi + \cos\theta \diff \psi\right)\,,\\[1mm]
 A^{3}\underset{r\to \infty}{\longrightarrow} \,& -\frac{\ii G_6 Q_3}{2\pi^2 r}\diff t_{\rm E} +\frac{G_6 J_-}{4\pi^2 r}\left(\diff \psi+\cos\theta \diff \phi\right)+\frac{G_6 J_+}{4\pi^2 r}\left(\diff\phi + \cos\theta \diff \psi\right)\,,
\end{aligned} 
\end{equation}
where $Q_1, Q_2, Q_3$ and $J_{\pm}$ are given in \eqref{eq:charges_STU}.
The asymptotic type IIB metric \eqref{eq:typeIIBmetric} can be conveniently described in terms of a new radial coordinate $\rho$, defined as
\begin{equation}\label{eq:rho}
r\,=\,\frac{\lambda^2}{4}\rho^2\,,\end{equation}
with
\begin{equation}\label{eq:lambda}
\lambda \,=\,\frac{2G_6}{\pi^2}\sqrt{Q_1 Q_2}\,=\,\sqrt{\frac{2G_6}{\pi^2}N_{\rm D1}N_{\rm D5}}\,,
\end{equation}
and it reads
\begin{equation}\label{eq:asymptotictypeIIBmetric}
\begin{aligned}
 \diff s^2_{\rm IIB} \,\underset{\rho\to \infty}{\longrightarrow} \,&\,   \lambda\left\{\frac{\diff \rho^2}{\rho^2}+\rho^2\left(\diff t_{\rm E}^2+\diff y^2\right)+\frac{1}{4}\left[\left(\diff \psi+\AL+\cos\theta \left(\diff \phi+\AR\right)\right)^2\right.\right.\\[1mm]
 \,&\,\quad \left.\left.+\,\diff \theta^2+ \sin^2\theta\left(\diff\phi+\AR\right)^2 \right]\right\}+ \sqrt{\frac{N_{\rm D1}}{v N_{\rm D5}}}\, \diff s^2_{M_4}\, ,
\end{aligned}
\end{equation}
where 
\begin{equation}\label{eq:3dvectors}
{\AL}\,=\,-\frac{2J_-}{N_{\rm D1} N_{\rm D5}}\left(\diff y+\ii\,\diff t_{\rm E}\right)\,, \hspace{1cm} {\AR}\,=\,-\frac{2J_+}{N_{\rm D1} N_{\rm D5}}\left(\diff y-\ii\,\diff t_{\rm E}\right)\,. 
\end{equation}
Hence, asymptotically the space is an $S^3$ fibration over Euclidean AdS$_{3}$. Both the AdS$_3$ and the $S^3$ radii are equal to $\sqrt\lambda$. Then, in order to trust the supergravity approximation, we need to take $N_{\rm D1}, N_{\rm D5}\gg 1$.   

The Brown-Henneaux formula then provides the central charge of the dual CFT,\footnote{The Brown-Henneaux formula for the CFT$_2$ central charge is $\cc = \frac{3\ell}{2G_3}$, where $\ell$ and $G_3$ are the AdS$_3$ radius and the 3d Newton's constant~\cite{Brown:1986nw}. Using $\ell = \lambda^{1/2}$ for the AdS$_3$ radius together with the relation $1/G_3 =  {\rm Vol}_{S^3}/ G_6 =  2\pi^2\lambda^{3/2}/G_6$ between the 6d and 3d Newton's constants, we obtain
$\cc= \frac{3\pi^2\lambda^{2}}{G_6} = \frac{12 G_6}{\pi^2}Q_1Q_2 = 6 N_{\rm D1} N_{\rm D5}$,
where in the second equality we plugged in $\lambda = \frac{2}{\pi^2}G_6\sqrt{Q_1Q_2}$, and in the third we used~\eqref{eq:integer_charges}, \eqref{eq:Newtons_const} in order to convert to integer units. 
As we discuss in some more detail at the end of this section, this expression for the central charge is valid at leading order and receives $\frac{1}{N_{\rm D1}N_{\rm D5}}$ corrections when $M_4=\Kt$.}
\be\label{eq:cisN1N5}
\cc  
\,=\, 6  N_{\rm D1} N_{\rm D5}\,.
\ee
The fact that the AdS radius and the CFT central charge are usually held fixed in holography\footnote{We note {\it en passant} that it is sometimes useful to introduce a chemical potential for the central charge, see~\cite{Dijkgraaf:1996xw} for a well-known example in the present context. The residues of the resulting  partition function have recently been connected with bulk saddles in \cite{Lee:2025veh}.}
is the reason why in this section we are switching to an ensemble of fixed D1- and D5-brane charges.

The conformal boundary of AdS$_{3}$ is a two-dimensional torus, which can be conveniently parametrized using complex coordinates,
\begin{equation}
w\,=\, y +\ii t_{\rm E}\,, \hspace{1.5cm} \bar w\,=\, y-\ii t_{\rm E}\,,
\end{equation}
periodically identified as
\begin{equation}\label{eq:identifications_w}
w\sim w+2\pi \sim w +2\pi\tau\, ,\hspace{1.5cm} {\bar w}\sim {\bar w}+2\pi \sim {\bar w} +2\pi{\bar\tau}\, .
\end{equation}
Then, eq.~\eqref{eq:periodicidentifications} leads to the identifications
\begin{equation}\label{eq:tau}
\tau\,=\,\frac{\varphi^3}{2\pi\ii}\, , \hspace{1.5cm} \bar \tau\,=\,\frac{\varphi^3+2\beta}{2\pi\ii}\,.
\end{equation}
For later convenience, we also introduce the combinations
\begin{equation}\label{eq:tau2}
\tau_1\,=\,\frac{\tau + \bar \tau}{2}\,=\,\frac{\varphi^3+\beta}{2\pi\ii}\,, \hspace{1.5cm} \tau_2\,=\,\frac{\tau-\bar\tau}{2\ii}\,=\,\frac{\beta}{2\pi}\,,
\end{equation}
noting that we do not require them to be necessarily real.\footnote{Expressing $\varphi^3 = \beta(\Phi -1)$, where $\Phi$ is the usual electrostatic potential introduced in black hole thermodynamics, we obtain the more symmetric expressions $\tau = \frac{\beta}{2\pi \ii}(\Phi -1)$, $\bar\tau = \frac{\beta}{2\pi \ii}(\Phi +1)$. It follows that $\tau_1 = \frac{\beta\Phi}{2\pi\ii}$ and $\tau_2 = \frac{\beta}{2\pi}$.
When $\beta$ is real and $\Phi$ is purely imaginary, $\tau$ and $\bar \tau$ are complex conjugate variables (equivalently, $\tau_1,\tau_2$ are real), otherwise they are not. From these observations, one also infers that $w,\bar w$ generically are not complex conjugate coordinates (this also follows from $t_{\rm E}$ and $y$ being generically not real coordinates, as it is manifest from \eqref{eq:periodicidentifications}). However, one can always make a change of coordinates such that the new complex coordinates are complex conjugate to each other. In these new coordinates, the metric has complex components.}
Moreover, we observe that the gauge fields \eqref{eq:3dvectors} are provided in a gauge such that ${\AL}$ ($\AR$) only has legs along $w$ ($\bar w$). In addition, such component depends on $J_{-}$ ($J_{+}$), which means that it should not be part of our set of boundary conditions, since we are not keeping the angular momenta fixed. Indeed, it is well known \cite{Dijkgraaf:2000fq, Kraus:2006nb,  Kraus:2006wn} that one component of the gauge fields---which in our conventions is ${\AL}_{w}$ (${\AR}_{\bar w}$)---corresponds to the VEV of the current in the dual field theory, while the other, $\AL_{\bar w}$ (${\AR}_{w}$), corresponds to the background field. Thus, only the latter quantity is part of our boundary conditions. 
 In order to see the components specifying the boundary conditions from the asymptotic form of the solution, we must introduce angular coordinates,
\begin{equation}
\psi'\,=\,\psi-\frac{\ii\omega_- }{\beta}\, t_{\rm E}\,, \hspace{1cm} \phi'\,=\,\phi-\frac{2\pi }{\beta}\, t_{\rm E}\,,
\end{equation}
which do not shift when going around the thermal circle. This is equivalent to performing the gauge transformations,
\begin{equation}
{\AL'}\,=\, {\AL} +\frac{\ii\omega_- }{\beta}\,\diff t_{\rm E}\,, \hspace{1cm} {\AR'}\,=\, {\AR} +\frac{2\pi}{\beta}\,\diff t_{\rm E}\,, 
\end{equation}
which leads to
\begin{equation}\label{eq:3dvectors2}
{\AL'}\,=\,-\frac{\omega_-}{2\beta}\,\diff {\bar w} +\left(\frac{\omega_-}{2\beta}-\frac{2J_-}{N_{\rm D1} N_{\rm D5}}\right)\diff{w}\,, \hspace{10mm} {\AR'}\,=\,-\frac{\pi \ii}{ \beta}\,\diff {w} + \left(\frac{\pi \ii}{\beta}-\frac{2J_+}{N_{\rm D1} N_{\rm D5}}\right)\diff {\bar w}\,.
\end{equation}
Finally, in order to facilitate a comparison with the dual field theory in section~\ref{sec:matchFareyTail}, we introduce 
\begin{equation}\label{eq:defomega}
\omega\,\equiv\,\ii \tau_2\,\AL'_{\bar w}  \,, \hspace{1cm}\tilde \omega\,\equiv\, \ii\tau_2\,\AR'_w\,,
\end{equation}
so that, from \eqref{eq:3dvectors2}, we get the following map with the angular potentials introduced in section~\ref{sec:multichargesaddles} in the asymptotically $S^1\times \mathbb{R}^4$ setup,
\begin{equation}\label{eq:omega_firsttime}
\omega\,=\,\frac{\omega_-}{4\pi\ii}\,, \hspace{1cm} \tilde \omega\,=\,\frac{\omega_+}{4\pi\ii}\,=\,\frac{1}{2}\,.
\end{equation}
In addition, since the partition function \eqref{eq:gravitationalindex} depends on $\omega_2$, it is useful to introduce
\begin{equation}\label{eq:defz}
z\,=\,\omega-\frac{1}{2}\,=\,-\frac{\omega_2}{2\pi\ii}\,.
\end{equation}
From now on we will adopt an AdS$_3/$CFT$_2$ notation and trade the thermodynamic potentials $\beta, \varphi^3, \omega_2$ for $\tau, \bar \tau, z$ using the relations above.

The analysis above shows that the decoupling limit of the solutions presented in section~\ref{sec:multichargesaddles} satisfies the requirements for contributing, \emph{a priori}, to a type IIB path integral with AdS${}_3\times S^3 \times M_4$ boundary conditions. If ${\cal X}$ denotes a generic field or state over which we sum in the path integral, we have that 
\begin{equation}
\begin{aligned}
{\cal X}\left(t_{\rm E}, y , \psi, \phi\right)\sim& \,(-1)^\FF{\cal X} \left(t_{\rm E} +2\pi\tau_2, y+2\pi\tau_1, \psi-2\pi \left(2z+1\right), \phi+2\pi\right)\\[1mm]
\sim& \, {\cal X}\left(t_{\rm E}, y+2\pi, \psi, \phi\right)\sim \,{\cal X}\left(t_{\rm E}, y, \psi +4\pi, \phi\right)\\[1mm]
\sim&\, (-1)^{\FF}{\cal X}\left(t_{\rm E}, y, \psi-2\pi, \phi+2\pi\right)\,.
\end{aligned}
\end{equation}
Combining the first and last, we get that fermions are periodic around the circle\footnote{More generally, we have that fermions are periodic around the circles parametrized by ${\cal X}\left(t_{\rm E}, y , \psi, \phi\right)\sim \,{\cal X} (t_{\rm E} +2\pi\tau_2, y+2\pi\tau_1, \psi-4\pi z +2\pi \left(n_2-n_1-1\right) , \phi+2\pi\left(1+n_1+n_2\right))$, if $n_1+n_2\in 2{\mathbb Z}+1$. Consistently with the regularity analysis in appendix~\ref{app:10Dregularity}, these circles never contract in the geometry. }
\begin{equation}
{\cal X}\left(t_{\rm E}, y , \psi, \phi\right)\sim \,{\cal X} (t_{\rm E} +2\pi\tau_2, y+2\pi\tau_1, \psi-4\pi z, \phi)\, .
\end{equation}
This precisely realizes the $(-1)^\FF$ insertion in the Hamiltonian representation of the gravitational index of type IIB on AdS$_3\times S^3 \times M_4$:
\begin{equation}\label{eq:IIBindex}
\begin{aligned}
Z \left(z, \tau\right)\,&=\,{\rm Tr}\, (-1)^{\FF} \rme^{-2\pi\tau_2 E+2\pi\ii \tau_1Q_3+2\pi \ii z\left(J_1-J_2\right)}\\[1mm]
\,&=\,{\rm Tr}\, (-1)^{\FF} \rme^{\pi\ii (\tau-\bar\tau) \{{\cal Q, \bar {\cal Q}}\}+2\pi\ii \tau Q_3+2\pi \ii z\left(J_1-J_2\right)}\,,
\end{aligned}
\end{equation}
where the trace is taken in a sector of fixed $Q_1,Q_2$ and the rewriting in the second line uses the superalgebra relations~\eqref{eq:5Dsuperalgebra} after the decoupling limit,
\begin{equation}\label{eq:3Dsuperalgebra}
\{{\cal Q, \bar {\cal Q}}\}\,=\, E- Q_3\,, \hspace{1cm} [J_i, {\cal Q}]\,=\,\frac{1}{2} {\cal Q}\,, \hspace{5mm} i=1, 2\,, \hspace{1cm} \left[Q_I, {\cal Q}\right]\,=\,0\,.
\end{equation}
This implies that the partition function only depends on $\tau, z$, and not on $\bar \tau$.

Formally, the gravitational index in the mixed ensemble of fixed $Q_1,Q_2$, that we will denote simply as $Z \left(\tau, z\right)$,
is related to the gravitational index $\hatZ\left(\varphi^1,\varphi^2,\varphi^3,\omega_2\right)$ where all chemical potentials are fixed---defined in \eqref{eq:gravitationalindex}---through the Laplace transform
\begin{equation}
    Z \left(\tau, z\right) \,=\,  \int\diff\varphi^1\int\diff\varphi^2 \,{\rm e}^{-\varphi^1 Q_1 - \varphi^2 Q_2}\,\hatZ\left(\varphi^1,\varphi^2,\varphi^3 = 2\pi\ii\tau, \omega_2 = -2\pi\ii z \right).
\end{equation} 
The corresponding relation between the on-shell action $I$ in the mixed ensemble and the on-shell action $\hatI$ in the grand-canonical ensemble, which represent saddle-point contributions to the respective partition functions, is given by the Legendre transform 
\begin{equation}
\label{eq:changeensemble}
    I \,=\, {\rm ext}_{(\varphi^1,\varphi^2)}\left(\hatI + \varphi^1 Q_1 + \varphi^2 Q_2\right)\,,
\end{equation}
 where ${\rm ext}_{(\varphi^1,\varphi^2)}$ denotes extremization over $\varphi^{1},\varphi^{2}$. 

We have written the gravitational index in the variables that will naturally appear in the dual CFT. 
As we will see, the partition function $Z(\tau,z)$ in \eqref{eq:IIBindex} can be identified with the elliptic genus of the dual D1-D5 CFT. In the next sections we will focus on $M_4 = \Kt$. Our results are also relevant for $M_4 = T^4$; although in this case the elliptic genus vanishes, it can be modified to a non-vanishing helicity supertrace~\cite{Maldacena:1999bp}. Given this, our next step is to evaluate saddle-point contributions to \eqref{eq:IIBindex} via their on-shell action, with the general goal of matching the corresponding expression in the dual CFT. Such holographic match, however, will be postponed until section~\ref{sec:matchFareyTail}, after we will have reviewed the relevant aspects of the elliptic genus and its Farey-tail expansion in section~\ref{sec:CFT2}. Now, we proceed with our discussion on the gravity side, focusing on the specific class of saddles with two  
centers, which will be those relevant for the match with the CFT.

\subsection{Decoupling limit of two-center saddles}\label{sec:decoupling2center}

Although the decoupling limit can be taken for all the solutions constructed in section~\ref{sec:multichargesaddles}, we now focus on the two-center class and provide additional details about their geometry after the decoupling limit. This generalizes the very recent analyses of \cite{Larsen:2026sav,Nanda:2026mbp,Georgescu:2026uhv}, which considered the saddle associated with the BMPV black hole (corresponding, in our notation, to the choice of integers $n=1$ and $p=q^1=q^2=q^3=0$).
 Here, we  study the decoupling limit of the full family of two-center solutions introduced in section~\ref{sec:multichargesaddles}. In section~\ref{sec:matchFareyTail}, we will argue that these solutions provide an infinite family of gravitational saddles contributing to the D1-D5 elliptic genus.

\paragraph{Local form of the solutions.} For two-center solutions, the geometry after the decoupling limit is locally AdS$_{3}\times S^3\times M_4$. This can be seen explicitly by introducing new coordinates $(\varrho, \vartheta)$ related to $(r,\theta)$ by\footnote{For two-center solutions, we can consider a region in parameter space in which $\delta$ is real and positive and $\varrho_+^2>\varrho^2_-$, so that the quantity inside the square root is always positive in the solution, which caps off at $\varrho=\varrho_+$.}
\begin{equation}
r^2\,=\,\frac{\lambda^4}{16}\left(\varrho^2-\varrho^2_+\right)\left(\varrho^2-\varrho^2_-\right)+\frac{\delta^2}{4}\cos^2\vartheta\,, \hspace{5mm} r\cos \theta\,=\,\cos \vartheta \sqrt{\frac{\lambda^4}{16}\left(\varrho^2-\varrho^2_+\right)\left(\varrho^2-\varrho^2_-\right)+\frac{\delta^2}{4}}\,,
\end{equation}
where $\lambda$ is given by \eqref{eq:lambda} and
\begin{equation}\label{eq:varrho_pm}
\varrho_{\pm}\,=\,\mp \frac{1}{4\kk^3_N} -\frac{h_N^2\left(1-h_N\right)^2\delta}{4\kk^1_N\kk^2_N\kk^3_N }\,.
\end{equation}
 In terms of these coordinates, the decoupled geometry reads
\begin{equation}\label{eq:decoupled_geometry}
\begin{aligned}
 \diff s^2_{\rm IIB} \,&=\,   \lambda \,\diff s^2_3+\frac{\lambda}{4}\left[\left(\diff \psi+\AL+\cos\vartheta \left(\diff \phi+\AR\right)\right)^2
+\,\diff \vartheta^2+ \sin^2\vartheta\left(\diff\phi+\AR\right)^2 \right] \\[1mm]
\,&\,+ \sqrt{\frac{N_{\rm D1}}{v N_{\rm D5}}}\, \diff s^2_{M_4}\, ,
 \end{aligned}
 \end{equation}
 where $\AL$ and $\AR$ are again the flat connections provided in \eqref{eq:3dvectors}, and
 \begin{equation}\label{eq:3dgeometry}
 \diff s^2_3\,=\,\frac{\left(\varrho^2-\varrho^2_+\right)\left(\varrho^2-\varrho^2_-\right)}{\varrho^2}\,\diff t_{\rm E}^2+\frac{\varrho^2 \diff \varrho^2}{\left(\varrho^2-\varrho^2_+\right)\left(\varrho^2-\varrho^2_-\right)}+\varrho^2\left(\diff y-\frac{\ii \varrho_+\varrho_-}{\varrho^2}\diff t_{\rm E}\right)^2\,. 
 \end{equation}
This three-dimensional metric is the familiar line element describing the Euclidean version of a non-extremal BTZ black hole~\cite{Banados:1992wn,Banados:1992gq}, however here we have additional identifications that will be described momentarily.

\paragraph{Thermodynamic variables and discrete data.} To begin with, one would like to find the dictionary between the parameters of the solution and the thermodynamic variables and discrete data. First, since we keep $N_{\rm D1}$ and $N_{\rm D5}$ fixed, we can use the first two expressions in \eqref{eq:charges_STU} to obtain that
\begin{equation}\label{eq:k1k2}
\kk^1_N\,=\, \frac{g_s\alpha' N_{\rm D1}}{4v} \frac{h_N\left(1-h_N\right)}{\kk^3_N}, \hspace{1cm} \kk^2_N\,=\,\frac{g_s\alpha'N_{\rm D5}}{4} \frac{h_N\left(1-h_N\right)}{\kk^3_N} .
\end{equation}
Next, we can use \eqref{eq:regularitycondIIB} and \eqref{eq:regularitycondIIB2} to obtain $\tau$ and $\bar \tau$ in terms of the parameters, yielding
\begin{equation}\label{eq:tau_2csaddles}
n\tau -q^3 \,=\,-2\ii \kk^3_N\,=\,\frac{\ii}{\varrho_+-\varrho_-} , \hspace{1cm} n{\bar\tau} -q^3\,=\, \frac{-\ii}{\varrho_++\varrho_-}\, ,
\end{equation}
or, equivalently via \eqref{eq:tau},
\begin{equation}
\frac{n\beta}{2\pi}\,=\,\frac{\varrho_+}{\varrho_+^2-\varrho_-^2}\,, \hspace{1cm}\frac{n}{2\pi}\left(\varphi^3+\beta\right)\,=\,-\frac{\varrho_-}{\varrho_+^2-\varrho_-^2}+\ii q^3\,.
\end{equation}
Finally, $h_N$ is related to the boundary condition $z$ defined in \eqref{eq:defz} via \eqref{eq:regularitycondIIB}:
\begin{equation}\label{eq:h_N->z}
h_N\,=\,p-nz \,.
\end{equation}
In what follows, we can use these relations to trade the parameters for the thermodynamic variables $\tau, \bar \tau, z$ and discrete data $q^3, p$. In particular, making use of them in the expressions for the momentum charge and angular momenta \eqref{eq:charges_STU}, we obtain
\begin{equation}\label{eq:charges_2c}
\begin{aligned}
N_{\rm P}\,&=\,N_{\rm D1} N_{\rm D5}\,\frac{ \left(n z-p+1\right) \left(nz-p\right)}{ \left(n \tau -q^3\right)^2}\,,\\[1mm]
J_+\,&=\,\frac{N_{\rm D1} N_{\rm D5}}{2}\, \frac{1}{n\bar \tau-q^3} \,, \hspace{1cm} J_-\,=\,-\frac{N_{\rm D1} N_{\rm D5}}{2}\frac{2nz+1-2p}{n\tau-q^3}\,,
\end{aligned}
\end{equation}
which further implies
\begin{equation}
\AL\,=\,\frac{2nz+1-2p}{n\tau-q^3}\,\diff w\,, \hspace{1cm} \AR\,=\,-\frac{\diff {\bar w}}{n{\bar \tau}-q^3}\,,
\end{equation}
after plugging \eqref{eq:charges_2c} in \eqref{eq:3dvectors}.

In order to describe the global properties of the manifold, we consider the $n\neq0$ and $n=0$ cases separately, starting with the first.

\paragraph{Supersymmetric orbifolds of BTZ $\boldsymbol{\times \, \, S^3}$.} From the general regularity analysis performed in appendix~\ref{app:10Dregularity}, we can easily extract the U(1) isometry that shrinks in the geometry \eqref{eq:decoupled_geometry}. This is generated by the following Killing vector
\begin{align}\label{eq:xi_orbifold_general}
\xi\,&=\,n\left(\frac{\beta}{2\pi}\partial_{t_{\rm E}}+\frac{\varphi^3+\beta}{2\pi\ii}\partial_{y}-\frac{\omega_-}{2\pi\ii}\partial_{\psi}+\partial_{\phi}\right) - q^3 \partial_y + p \left(\partial_{\phi}+\partial_{\psi}\right) + \left(1-n-p\right)\left(\partial_{\phi}-\partial_{\psi}\right) \nn\\[1mm]
\,&=\, n\tau_2\,\partial_{t_{\rm E}}+\left(n\tau_1-q^3\right)\,\partial_y + \left(-2nz+2p-1\right)\partial_{\psi}+\partial_{\phi}\,,
\end{align}
whose norm vanishes at $\varrho=\varrho_+$. The rewriting in the second line follows after using \eqref{eq:tau2} and \eqref{eq:defz}.
Let us then proceed by introducing a set of coordinates adapted to $\xi$, namely
\begin{equation}
{\tilde t}_{\rm E}\,=\,\frac{{t}_{\rm E}}{n\tau_2}\,, \hspace{5mm} {\tilde y}\,=\, y-\frac{n\tau_1-q^3}{n\tau_2}t_{\rm E}\,, \hspace{5mm} \tilde\psi\,=\,\psi- \frac{-2nz+2p-1}{n\tau_2}t_{\rm E}\,,\hspace{5mm} \tilde \phi\,=\,\phi-\frac{t_{\rm E}}{n\tau_2}\,,
\end{equation}
so that 
\begin{equation}
\xi\,=\,\partial_{\,{\tilde t}_{\rm E}}\,.
\end{equation}
From the three-dimensional point of view, this change of coordinates induces a gauge transformation on $\AL$ and $\AR$, 
\begin{equation}
\begin{aligned}
\AL\ \to\ & \AL+\left(-2nz+2p-1\right)\diff {\tilde t}_{\rm E}\,=\,\frac{2nz+1-2p}{n\tau-q^3}\,\diff {\tilde y}\,, \\[1mm]
\AR\ \to \ & \AR+\diff {\tilde t}_{\rm E}\,=\,-\frac{\diff{\tilde y}}{n\bar\tau-q^3}\,,
\end{aligned}
\end{equation}
which become regular in the new gauge. This describes (BTZ $\times$ $S^3)/{\mathbb Z}_{|n|}$, with the orbifold action given by
\begin{equation}\label{eq:BTZxS3orbifolds}
\begin{aligned}
\left(\tilde t_{\rm E}, {\tilde y}, {\tilde\psi},\tilde \phi\right)\sim& \left(\tilde t_{\rm E}+\frac{2\pi}{n}, {\tilde y}+\frac{2\pi q^3}{n}, {\tilde\psi}-\frac{2\pi\left(2p-1\right)}{n},\tilde \phi-\frac{2\pi}{n}\right)\\[1mm]
\sim& \left(\tilde t_{\rm E}, {\tilde y}+2\pi, {\tilde\psi},\tilde \phi\right) \sim \left(\tilde t_{\rm E}, {\tilde y}, {\tilde\psi}+4\pi,\tilde \phi\right) \sim \left(\tilde t_{\rm E}, {\tilde y}, {\tilde\psi}-2\pi,\tilde \phi+2\pi\right)\,,
\end{aligned}
\end{equation}
 
 An equivalent way of understanding the orbifold action is by noticing that $\xi$ can be written as
\begin{equation}
\xi\,=\,{\tilde \tau}_2\,\partial_{t_{\rm E}}+{\tilde \tau}_1\partial_y  -\left(2\tilde z+1\right)\, \partial_{\psi}+\partial_{\phi}\,, 
\end{equation}
with 
\begin{equation}
{\tilde \tau}_2\,=\, n\tau_2\,, \hspace{1cm} {\tilde \tau}_1 \,=\,n\tau_1-q^3\,, \hspace{1cm} \tilde z\,=\, nz-p\, .
\end{equation}
In terms of the tilded variables, the identifications become
\begin{equation}
\begin{aligned}
\left(t_{\rm E}, y , \psi, \phi\right)&\sim \left(t_{\rm E}+\frac{2\pi{\tilde\tau}_2}{n}, y +\frac{2\pi\left({\tilde \tau}_1+q^3\right)}{n}, \psi-\frac{4\pi\left(\tilde z+p\right)}{n}, \phi\right)\sim  \left(t_{\rm E}, y +2\pi, \psi, \phi\right)\\[1mm]
&\sim \left(t_{\rm E}, y , \psi+4\pi, \phi\right)\sim \left(t_{\rm E}, y , \psi-2\pi, \phi+2\pi\right)\,,
\end{aligned}
\end{equation}
which shows that the solution is an orbifold of a BTZ black hole with temperature $2\pi{\tilde \tau}_2$ and velocities $\tilde \tau_1$ and $\tilde z$, respecting the boundary conditions needed to contribute to a partition function with chemical potentials $\tau=\tau_1+\ii \tau_2$ and $z$. This makes it evident that our construction here is the AdS$_3\times S^3$ 
version of the orbifolds discussed by \cite{Aharony:2021zkr} for AdS$_5\times S^5$ black holes (see also \cite{BenettiGenolini:2023rkq} for the analogous construction of black hole index saddles in AdS$_4\times S^7$). 

When $n$, $q^3$ are coprime, this construction gives the ${\rm SL}(2, {\mathbb Z})$ family of BTZ black holes found in \cite{Maldacena:1998bw}, and clarifies how it is compatible with supersymmetry. Specifically, we find that for supersymmetry to be preserved the orbifold must act non-trivially on $S^3$, analogously to what occurs in the asymptotic ${\rm AdS}_5\times S^5$ case~\cite{Aharony:2021zkr}.  When $n$ and $q^3$ have a common factor $k$, we obtain new saddles, where the ${\rm SL}(2, {\mathbb Z})$ family of BTZ black holes sits in a $\mathbb{Z}_k$ orbifold of ${\rm AdS}_3\times S^3$.
As we will see, this is in agreement with the Farey-tail expansion \cite{Dijkgraaf:2000fq}, although the action of the orbifolds \eqref{eq:BTZxS3orbifolds} was not discussed there.

From the orbifold action \eqref{eq:BTZxS3orbifolds} we see that it is freely acting if and only if ${\rm gcd}\left(n, q^3, p\right)\,=\,{\rm gcd}\left(n, q^3, p-1\right)\,=\,1$. When $n, q^3$ and $p$ (and/or $n, q^3$ and $p-1$) have common factors, the orbifold has fixed points at the poles of the horizon $S^3$. The singularity at the north (south) pole, corresponding to $\varrho\,=\,\varrho_+$, $\vartheta\,=\,0 $ (or $\varrho=\varrho_+$, $\vartheta\,=\,\pi$) is of degree gcd$(n, q^3,p)$ (or gcd($n, q^3, p-1$)).  
The fixed loci correspond to the torus formed by $S^1_y$ and by the $S^1$ in the horizon $S^3$ that does not collapse at that pole.

\paragraph{Horizonless orbifolds.} Let us now turn our attention to the two-center saddles with $n=0$. In section~\ref{sec:multichargesaddles}, we claimed that these correspond to the Euclidean version of the solutions of \cite{Giusto:2012yz}, whose decoupling limit yields an orbifold of AdS$_3\times S^3$. We would like to see this explicitly here. To this aim, we first note that when setting $n=0$ in \eqref{eq:tau_2csaddles}, one finds that
\begin{equation}
\varrho_+\,=\,0\,, \hspace{1cm} \varrho_-\,=\,\frac{\ii}{ q^3}\,,
\end{equation}
so that the three-dimensional metric \eqref{eq:3dgeometry} can be written as 
\begin{equation}
\diff s^2_3\,=\,\frac{(\korb\varrho)^2+1}{\korb^2}\, \diff t_{\rm E}^2+\frac{\korb^2\,\diff \varrho^2}{(\korb\varrho)^2+1} +\varrho^2\,{\diff y}^2\,, 
\end{equation}
where, to facilitate the comparison with \cite{Giusto:2012yz}, we have defined for these solutions
\begin{equation}\label{eq:q3_is_k}
\korb\equiv -q^3\, . 
\end{equation}
The charges of the solution can be simply obtained by setting $n=0$ in \eqref{eq:charges_2c}, which yields
\begin{equation}\label{eq:charges_soliton2}
N_{\rm P}\,=\, \frac{N_{\rm D1}N_{\rm D5}}{\korb}\frac{p(p-1)}{\korb}\,,\hspace{1cm} J_+\,=\,\frac{N_{\rm D1}N_{\rm D5}}{2\korb}\,, \hspace{1cm} J_-\,=\,\frac{N_{\rm D1}N_{\rm D5}}{2\korb}(2p-1)\, ,
\end{equation}
implying that the expression for the gauge fields boils down to
\begin{equation}\label{eq:gaugefields_soliton}
{\AL}\,=\,\frac{1-2p}{\korb}\,\diff{w}\,, \hspace{10mm} {\AR}\,=\,-\frac{\diff {\bar w}}{\korb}\,.
\end{equation}
Given this information, one has that the U(1) contracting in the decoupled geometry \eqref{eq:decoupled_geometry} is generated by the Killing vector (see eq.~\eqref{eq:rodvector6D})
\begin{equation}\label{eq:U(1)_soliton}
\xi\,=\,\korb\,\partial_y +\partial_{\phi}+\left(2p-1\right)\partial_{\psi}\, ,
\end{equation}
whose norm vanishes at $\varrho=0$. Therefore, the decoupled geometry in the $n=0$ case is Euclidean (${\rm AdS}_3\times S^3)/{\mathbb Z}_{|k|}$. As we did for the $n\neq0$ case before, we introduce coordinates adapted to the isometry \eqref{eq:U(1)_soliton}, 
\begin{equation}
\tilde y\,=\, \frac{y}{\korb}\,, \hspace{1cm} \tilde\psi \,=\, \psi+(1-2p)\,\frac{y}{\korb}\,, \hspace{1cm} \tilde \phi\,=\, \phi-\frac{y}{\korb}\,,
\end{equation}
so that $\xi = \partial_{\tilde y}$. The effect of this coordinate transformation is to cancel the $y$-component of the gauge fields in \eqref{eq:gaugefields_soliton}, at the expense of introducing twisted identifications for the new coordinates: 
\begin{equation}\label{eq:orbifold_soliton}
\left(\tilde y, \tilde \psi, \tilde \phi\right)\sim \left(\tilde y+\frac{2\pi}{\korb}, \tilde \psi+\frac{2\pi\left(1-2p\right)}{\korb}, \tilde \phi-\frac{2\pi}{\korb}\right)\sim \left(\tilde y, \tilde \psi+4\pi, \tilde \phi\right)\sim \left(\tilde y, \tilde \psi+2\pi, \tilde \phi+2\pi\right) \,.
\end{equation}
At this stage the analysis parallels that of \cite{Giusto:2012yz}. The orbifold only acts freely when gcd$(p, \korb)=1$ and gcd$(p-1, \korb)=1$. In this case, the hypersurface at $\varrho=0$ is topologically $S^1\times S^3$. In turn, when either gcd$(p, \korb)>1$ or gcd$(p-1, \korb)>1$, one has conical singularities at the poles of $S^3$. 

The possible values of $\korb, p$ are constrained by quantization conditions to be imposed on physical quantities.  
 A first set of conditions arises from evaluating $q^1$, $q^2$ in the horizonless solutions and demanding that the resulting expressions be integer. Combining \eqref{eq:regularitycondIIB2} with \eqref{eq:k1k2} and \eqref{eq:h_N->z} for $n=0$, we deduce that
\begin{equation}\label{eq:q1q2_soliton}
q^1\,=\, -N_{\rm D1}\, \frac{p(p-1)}{\korb} \,\in\, \mathbb Z\,,\hspace{1cm}q^2\,=\,-N_{\rm D5}\, \frac{p(p-1)}{\korb} \,\in \,\mathbb Z\,.
\end{equation}
We will discuss further quantization conditions, to be imposed on conserved charges, momentarily.

\paragraph{On-shell action at fixed D1-D5 charges.} The on-shell action of the two-center saddles prior to the decoupling limit was discussed in section~\ref{sec:2centresaddle}. It is straightforward to check that it remains finite in the decoupling limit, so 
the same expression provides the action of the asymptotically ${\rm AdS}_3\times S^3$ configurations. Therefore, in order to obtain the action at fixed D1-D5 charges, that we denote by $I$, we just need to take the Legendre transform \eqref{eq:changeensemble} of the grand-canonical action given in \eqref{eq:2centregeneral}. This yields
\begin{equation}
\label{eq:BHEG}
    I \,=\, 2\pi\ii\,\Psi - 2\pi\ii\,N_{\rm D1}N_{\rm D5}\,\frac{1 - \left(2n z -2p +1\right)^2}{4n\left(n\tau - q^3\right)}\,,
\end{equation}
\be\label{eq:phase_term}
\Psi \,=\, \frac{1}{n}\left[q^1 N_{\rm D5} + q^2 N_{\rm D1} + w_0^2(w_0^1)^2(n+p-1)   \,q^1 q^2 q^3 \right]\,,
\ee
where we used \eqref{eq:tau} and \eqref{eq:defz} to convert the potentials $\varphi^3,\omega_2$ into $\tau,z$, as well as \eqref{eq:integer_charges} with $\frac{2G_6 }{\pi^2}e^1 e^2=1$ and $e^3 =1$ to express $Q_1,Q_2$ in terms of the integers $N_{\rm D5},N_{\rm D1}$. The term $\Psi$ is independent of $\tau,z$ and gives rise to a pure phase in $\rme^{-I}$.
In particular, the Legendre transform of the horizonless soliton action \eqref{eq:horizonlessgeneral} yields
\begin{equation}
\label{eq:solitonEG}
    I_{n\to 0} \,=\, -2\pi \ii\,\frac{N_{\rm D1}N_{\rm D5}}{k} \left[\frac{p\left(p-1\right)}{k}\,\tau+ \left(2p-1\right)z  \right]\,.
\end{equation}

\vskip 3mm

\paragraph{Quantum corrections and charge quantization.} Although our analysis so far assumed classical two-derivative supergravity, it is not hard to include one-loop corrections from type IIB supergravity in our formulae for the conserved charges and on-shell action. Taking these corrections into account will allow us to impose  precise quantization conditions on the conserved charges.

We rely on the results of~\cite{Beccaria:2014qea,ArabiArdehali:2018mil}, where the correction we are after is obtained by evaluating the ${\rm AdS}_3\times S^3\times M_4$ vacuum energy, with $M_4 = T^4$ or ${\rm K3}$, and corresponds to a correction to the central charge of the dual CFT$_2$. The same correction is expected to appear as an overall correction to the on-shell action and charges of our solutions. While in the case of $T^4$ there are no corrections, in the case of ${\rm K3}$ one finds a non-trivial correction, which can be expressed as the shift
\be\label{eq:quantum_correction}
N_{\rm D1}N_{\rm D5} \ \rightarrow \ N_{\rm D1}N_{\rm D5} + 1 \, \equiv \, N\,.
\ee
Taking this into account, the central charge \eqref{eq:cisN1N5}
becomes $\cc 
\,=\, 6N$,
which precisely matches the microscopic calculation (based on the study of the instanton moduli space on $M_4$).
The modification \eqref{eq:quantum_correction} should be implemented in the charges and on-shell action of both the black hole and soliton saddles, given in~\eqref{eq:charges_2c},  \eqref{eq:charges_soliton2}, \eqref{eq:BHEG} and \eqref{eq:solitonEG}. The phase $\Psi$ is corrected in the same way,
\be\label{eq:correction_Psi}
\Psi \ \rightarrow \ \frac{N}{N_{\rm D1}N_{\rm D5}}\,\Psi\,.
\ee

Having obtained more accurate expressions, we can now impose suitable charge quantization conditions. These conditions arise in horizonless solutions, by demanding that they correspond to individual states in the D1-D5 CFT, which carry integer charges.
Imposing quantization of the angular momenta for the two-center solitons, given in \eqref{eq:charges_soliton2}, one finds the condition
\begin{equation}\label{eq:qcsol1}
\frac{N}{\korb}\,\in\, {\mathbb Z}\,,
\end{equation}
which is indeed assumed in the construction of the dual CFT states \cite{Giusto:2012yz}.\footnote{Although ref.~\cite{Giusto:2012yz} only considers the case $M_4 = T^4$, the construction extends to $M_4 = {\rm K}3$.}
The momentum charge in \eqref{eq:charges_soliton2} should also be quantized, namely we need
\be
\frac{N}{\korb}\frac{p(p-1)}{\korb} \,\in\,\mathbb{Z}\,.
\ee
In the case of $M_4 ={\rm K}3$, combining \eqref{eq:qcsol1} with~\eqref{eq:q1q2_soliton} implies
 ${\rm gcd}(k,N_{\rm D1}) = {\rm gcd}(k,N_{\rm D5}) = 1 $, and therefore
\be\label{eq:qcsol3}
\frac{p(p-1)}{\korb} \,\in\,\mathbb{Z}\qquad \text{for} \ M_4 = {\rm K}3\,.
\ee
Quantization of the momentum charge is then guaranteed. Note that since in this case either gcd$(p, \korb)>1$ or gcd$(p-1, \korb)>1$, the orbifold is always singular. In the case of $M_4 = T^4$, where $N = N_{\rm D1}N_{\rm D5}$ with no shift by 1, one only deduces from \eqref{eq:q1q2_soliton} the weaker condition
\begin{equation}\label{eq:qcsol2}
\frac{p(p-1)}{\korb} \,{\text {gcd}}(N_{\rm D1}, N_{\rm D5})\,\in\, {\mathbb Z} \qquad \text{for} \ M_4 = T^4\,.
\end{equation} 
Hence quantization of the momentum is automatically guaranteed only if ${\text {gcd}}(N_{\rm D1}, N_{\rm D5})=1$, otherwise it must be imposed independently~\cite{Giusto:2012yz}.

\paragraph{Summary.} 
In this section we have set the stage for studying the asymptotically flat saddles within AdS$_3$/CFT$_2$ holography upon taking the decoupling limit.
Focusing on two-center saddles, we have exhibited an infinite family of configurations possibly contributing to the gravitational index of type IIB string theory with Euclidean AdS$_3\times S^3\times M_4$ boundary conditions. The geometries are in fact locally AdS$_{3}\times S^3\times M_4$, however they differ by global identifications. In addition to the fixed charges $N_{\rm D1},N_{\rm D5}$ and the chemical potentials $\tau,z$, the saddles depend on the integers $n,p,q^1,q^2,q^3$, specifying this global information. The integers  $n,p,q^3$, that specify an orbifold action, are coupled to the chemical potentials and enter in the expressions for the charges, while $q^1,q^2$ only appear in a constant phase term in the on-shell action. We have found the on-shell action both of the horizonless saddles (namely, those with $n=0$) and the black hole saddles (i.e.\ those with $n\neq0$).
  In section~\ref{sec:matchFareyTail}, focusing on $M_4=\Kt$, we will match the action and establish a precise map between these saddles and those of the dual CFT$_2$ elliptic genus.

\section{The D1-D5 elliptic genus and its Farey-tail expansion}\label{sec:CFT2}

In this section, we review some basic features of the elliptic genus of the two-dimensional SCFT describing the low-energy limit of the system of D1-D5 branes wrapping $S^1\times {\rm K}3$, dual to type IIB string theory on ${\rm AdS}_3 \times S^3 \times {\rm K}3$. In particular, we recall the Farey-tail expansion of the elliptic genus~\cite{Dijkgraaf:2000fq}. This is an exact rewriting of the D1-D5 elliptic genus as a sum over modular images of a finite set of contributions, known as polar states. The expansion suggests an interpretation of the different terms in the sum as dual asymptotically Euclidean AdS$_3$ geometries. 

\subsection{The elliptic genus}

The elliptic genus of a two-dimensional SCFT is its supersymmetric index on the torus \cite{Schellekens:1986xh,Witten:1986bf}.
Although we are interested in the  SCFT describing the D1-D5 system on ${\rm K}3\times S^1$, which has $(4,4)$ supersymmetry, in this section we will work more generally and discuss the elliptic genus of a generic $(2,2)$ SCFT, following~\cite{Moore:2004fg}. 

The $(2,2)$ superconformal algebra is given by two copies of the ${\cal N}\,=\,2$ super-Virasoro algebra, ${\rm Vir}^{{\cal N}=2}_{\rm left} \bigoplus {\rm Vir}^{{\cal N}=2}_{\rm right}$, with the left-moving copy ${\rm Vir}^{{\cal N}=2}_{\rm left} $ being generated by $L_n, G^{\pm}_r, J_n$, and the right-moving copy ${\rm Vir}^{{\cal N}=2}_{\rm right} $ being generated by ${\tilde L}_n, {\tilde G}^{\pm}_r, {\tilde J}_n$. Here, $L_n,\tilde L_n$, $n\in\mathbb Z$, are Virasoro generators,  $J_n,\tilde J_n$ are U(1) R-current generators, and $G^\pm_r,\, \tilde G^\pm_r$ are supercurrent generators, with  $r,s \in \mathbb Z$ in the Ramond (R) sector and $r,s \in \frac{1}{2}+\mathbb Z$ in the Neveu-Schwarz (NS) sector of the theory on the cylinder (or on the torus).  We provide the relations obeyed by the left-moving generators, those for the right-moving ones being similar:
\begin{equation}
\begin{aligned}
\label{eq:GG}
 [L_n, L_m]&\,=\, \left(n-m\right)L_{n+m}+\frac{\cL}{12}n\left(n^2-1\right)\delta_{n+m, 0}\,, \\[1mm]
 \{G^\pm_r, G^\mp_s\}&\,=\, 2L_{r+s}+\left(r-s\right)J_{r+s}+\frac{\cL}{12}\left(4r^2-1\right)\delta_{r+s, 0}\,, \\[1mm]
 [J_n, J_m]&\,=\, \frac{\cL}{3}n\,\delta_{n+m, 0}\,, \\[1mm]
 \left[L_n, G^\pm_r\right]&\,=\,\left(\frac{n}{2}-r\right)G^{\pm}_{n+r}\,, \\[1mm]
  \left[J_n, G^\pm_r\right]&\,=\,\pm G^{\pm}_{n+r}\,, \\[1mm]
   \left[L_n, J_m\right]&\,=\,-m J_{n+m}\,,
\end{aligned}
\end{equation}
where $\cc$ denotes the central charge.\footnote{Since we are considering a non-chiral theory (with $(2,2)$ or $(4,4)$ supersymmetry), the left- and right-moving central charges are the same. The label $n$ appearing in \eqref{eq:GG} is unrelated to the integer $n$ introduced in the previous section.}
It will also be useful to recall that the ${\cal N}=2$ superconformal algebra admits a family of \emph{spectral flow} isomorphisms, defined as 
\begin{equation}\label{eq:spectral_flow}
G^{\pm}_{n\pm a}\to G^{\pm}_{n\pm (a+\eta)}\,, \hspace{1cm} L_n \to L_n +\eta J_n +\frac{\cL}{6}\,\eta^2\,\delta_{n, 0}\,, \hspace{1cm} J_n \to J_n + \frac{\cL}{3}\,\eta\,\delta_{n, 0}\,,
\end{equation} 
where the parameter $\eta$ may be integer or half-integer. 
Half-integer values of $\eta$ connect the R and NS sectors, while integer values act within the same sector. 

 The elliptic genus is defined as the following trace in the Hilbert space of the theory with R boundary conditions for both left- and right-movers,
\begin{equation}\label{eq:ellipticgenus}
 \chi_{\rm RR}\left(\tau, z\right) 
 \,=\, {\rm Tr}_{\rm RR} \,(-1)^\FF \, \rme^{2\pi \ii \tau \left(L_0-\frac{\cL}{24}\right)-2\pi \ii {\bar \tau}\left({\tilde L}_0-\frac{\cL}{24}\right)+2\pi \ii z J_0 }\,,
\end{equation}
where $\FF$ is the fermion number operator and $\tau$, $\bar \tau$, $z$ are complex parameters.
The standard Witten index argument~\cite{Witten:1982df}, together with the superalgebra relation $\frac12\{{\tilde G}^+_0, {\tilde G}^-_0\}\,=\,{\tilde L}_0-\tfrac{\cL}{24}$, imply that the elliptic genus only receives contributions from states annihilated by both ${\tilde G}^{\pm}_0$, namely from the right-moving ground states, 
 and is therefore independent of $\bar\tau$. All states in the left-moving sector contribute instead.

Recalling that the fermion number can be realized as 
\be
(-1)^{\FF} = \rme^{\pi \ii ({J}_0- \tilde J_0)}\,,
\ee 
the elliptic genus can be obtained starting from the more general non-supersymmetric  partition function 
\begin{equation}\label{eq:gen_part_fct_3}
 Z_{\rm RR}\left(\tau, \bar \tau, \omega, {\tilde \omega}\right)\,=\, {\rm Tr}_{\rm RR} \, \rme^{2\pi \ii \tau \left(L_0-\frac{\cL}{24}\right)-2\pi \ii {\bar \tau}\left({\tilde L}_0-\frac{\cL}{24}\right)+2\pi \ii \omega J_0 -2\pi\ii {\tilde \omega} {\tilde J}_0}\,,
\end{equation}
depending on the four variables $\tau,
\bar\tau,\omega,\tilde\omega$, and imposing the condition 
\begin{equation}
\label{eq:omegatildeomega}
{\tilde \omega}\,=\,\frac{1}{2}\,,\qquad \qquad \omega = z + \frac{1}{2}\,,
\end{equation}
where the second is simply a convenient redefinition. In other words, the elliptic genus is given by
\be\label{eq:ell_gen_from_ZRR}
\chi_{\rm RR}\left(\tau, z\right)\,=\, Z_{\rm RR}\left(\tau, \bar\tau, z + \tfrac{1}{2}, \tfrac{1}{2}\right).
\ee
Note that since all  charge operators appearing in \eqref{eq:gen_part_fct_3} have integer eigenvalues,  each of the variables $\tau,
\bar\tau,\omega,\tilde\omega$ is  defined modulo shifts by 1. The condition \eqref{eq:omegatildeomega} is also imposed modulo 1.\footnote{While periodicity under integer shifts of the potentials is clear for the SCFT partition function considered here, it is less apparent when the gravitational index is considered, since this symmetry is only recovered by summing over an infinite family of saddles; see e.g.~\cite{Goker:2026tct} for a recent discussion. Nevertheless, we will use the same notation to indicate the SCFT and the gravitational chemical potentials.}

Recalling that the Hamiltonian and the momentum on the cylinder are given by
\begin{equation}\label{eq:HP_CFT}
 H\,=\, L_0 + {\tilde L}_0-\frac{\cL}{12}\,,\qquad \qquad
 P\,=\, L_0-{\tilde L}_0\,,
\end{equation}
 the partition function \eqref{eq:gen_part_fct_3} can also be expressed as
\begin{equation}\label{eq:ZRR_grandcanonical}
 Z_{\rm RR}\left(\tau, \bar \tau, \omega, {\tilde \omega}\right)\,=\, {\rm Tr}_{\rm RR} \, \rme^{-2\pi \tau_2 \left(  H - \ii\frac{\tau_1}{\tau_2} P - \ii \frac{\omega}{\tau_2} J_0 + \ii \frac{\tilde \omega}{\tau_2} {\tilde J}_0\right)} \,,
\end{equation}
where we used $\tau\,=\, \tau_1+ \ii \tau_2$ and $\bar \tau = \tau_1-\ii \tau_2$. 
Eq.~\eqref{eq:ZRR_grandcanonical} makes it manifest that the elliptic genus can be obtained by imposing the condition \eqref{eq:omegatildeomega} on a standard grand-canonical partition function, with $2\pi \tau_2$ playing the role of inverse temperature and the variables $\ii\frac{\tau_1}{\tau_2}$, $\ii\frac{\omega}{\tau_2}$ and $-\ii\frac{\tilde\omega}{\tau_2}$  being interpreted as standard chemical potentials conjugate to $P$, $J_0$ and $\tilde J_0$, respectively.

The expressions \eqref{eq:gen_part_fct_3}--\eqref{eq:ZRR_grandcanonical} lead to a convenient path integral representation for the elliptic genus. This is the Euclidean path integral defined by compactifying the cylinder to a torus with modular parameter $\tau$, and fixing suitable boundary conditions  together with appropriate background gauge fields coupling to the left and right-moving R-currents. The torus is parameterized by the complex coordinate $w\,=\, y +\ii t_{\rm E}$, satisfying the identifications $w  \sim w+2\pi \sim w + 2\pi \tau $.
The requirement of being in the RR sector means that all fields (either bosonic or fermionic) are periodic when going once around the spatial cycle of the torus, 
\begin{equation}\label{identif_ell_genus}
 {\cal X}_{\rm CFT}(w+2\pi) \,\sim\,  {\cal X}_{\rm CFT}(w)\,\quad \qquad \text{(same for right-movers, depending on $\bar w$)}\,.
\end{equation}
On the other hand, the fact that there is no explicit $(-1)^{\FF}$ insertion in \eqref{eq:gen_part_fct_3} means that we are choosing identifications around the thermal circle where bosons are periodic while fermions are antiperiodic, namely
\begin{equation}\label{identif_ell_genus_2}
{\cal X}_{\rm CFT}(w+2\pi \tau) \,\sim\, (-1)^{\FF}{\cal X}_{\rm CFT}(w)\,\quad \qquad \text{(similar for right-movers, with $\tau\rightarrow\bar\tau$)}\,.
\end{equation}
In addition, we have to turn on the chemical potentials \eqref{eq:omegatildeomega} for the R-charges. This can be done by considering the following background gauge field components $\AL'_{\bar w}$, $\AR'_w$ that couple to the left- and right-moving R-currents, respectively, 
\begin{equation}\label{eq:bckgnd_gauge_fields}
\AL'_{\bar w}\,=\,-\frac{ \ii\omega}{ \tau_2} \,=\, -\frac{ \ii }{ \tau_2}\left(z+\frac12\right)  \,, \qquad\qquad \AR'_w\,=\,-\frac{ \ii\tilde\omega}{ \tau_2}\,=\,-\frac{\ii}{ 2\tau_2} \,.
\end{equation}
These gauge fields are just the boundary gauge fields introduced in the previous section, cf.\ eq.~\eqref{eq:3dvectors2}. We are thus adopting the same notation.

We stress that different but equivalent choices are possible. Indeed, by implementing suitable large R-symmetry transformations, we can modify the background gauge fields while introducing an R-symmetry twist in the boundary conditions for charged fields. In this way, we may arrange for periodic boundary conditions for all fields around both cycles of the torus,
\begin{equation}
{\cal X}_{\rm CFT}(w+2\pi \tau) \,\sim\, {\cal X}_{\rm CFT}(w) \,\sim\,  {\cal X}_{\rm CFT}(w+2\pi)\, \qquad \text{(similar for right-movers, with $\tau\rightarrow\bar\tau$)}\,,
\end{equation}
with the background gauge fields now being $
{\AL}_{\bar w}=-\ii\frac{ z}{\tau_2} \,$, $ {\AR}_w=0$. This choice directly matches the trace representation \eqref{eq:ellipticgenus}. Alternatively, by a further R-symmetry transformation we may set ${\AL}_{\bar w}= {\AR}_w=0$ and encode the full R-symmetry chemical potentials in twisted boundary conditions for the fields,
\begin{equation}
{\cal X}_{\rm CFT}(w+2\pi \tau) \,\sim\, \rme^{2\pi  \ii  z j}{\cal X}_{\rm CFT}(w)\,, \qquad\quad {\cal X}_{\rm CFT}(\bar w+2\pi \bar\tau) \,\sim\, {\cal X}_{\rm CFT}(\bar w)\,, 
\end{equation}
where $j$ is the charge of a left-moving field under $J_0$.
In the following we will refer to either one of these equivalent pictures.

\subsection{The Farey-tail expansion}

We now focus on the elliptic genus of the D1-D5 SCFT. Since this theory preserves $(4,4)$ supersymmetry, the U(1)$_{\rm left}$ and U(1)$_{\rm right}$ current algebras generated by $J_n$ and $\tilde J_n$ are enhanced to level $\nnn$ affine SU(2)$_{\rm left}$ and SU(2)$_{\rm right}$ current algebras, and the central charge is given by
\be
\cc = 6\nnn\,.
\ee
We will use $N$ instead of $\cc$ from now on. The generators $J_0$ and $\tilde J_0$ have integer eigenvalues in the R sector.

The D1-D5 elliptic genus can be computed using the symmetric product orbifold CFT and is explicitly known~\cite{Dijkgraaf:1996xw}.
In \cite{Dijkgraaf:2000fq}, it was found that it admits the following Farey-tail expansion   
\begin{align}\label{eq:Fareytail_exp}
        \chi_{\rm RR}\left(\tau,z\right) &= -2\pi  \ii\sum_{\left(c,d\right)}\sum_{\mu=-\nnn+1}^{\nnn} \sum_{0\leq m< \frac{\mu^2}{4\nnn} }\frac{\tilde c_{(\mu)}}{\left(c\tau + d\right)^3}\,{\rm exp}\left[2\pi \ii \left( m - \frac{\mu^2}{4\nnn}\right)\frac{a\tau+b}{c\tau+d} - 2\pi \ii \nnn\frac{cz^2}{c\tau + d}\right] \times
       \nn \\[1mm]
        &\qquad\qquad \times\Theta_{(\mu,\nnn)}\left(\frac{z}{c\tau+d}, \frac{a\tau+b}{c\tau+d}\right)\,.
    \end{align}
This follows from applying the Jacobi-Rademacher expansion formula to the Farey-tail transform of the elliptic genus, see~\cite{Moore:2004fg} for a review.\footnote{The original work of~\cite{Dijkgraaf:2000fq} applied a Jacobi-Rademacher expansion to the `Farey-tail transform' of the elliptic genus, while later formulations reconstructed directly the elliptic genus through a related expansion~\cite{Manschot:2007ha}. These two prescriptions 
yield the same result at the semiclassical level, which is our focus.}
The expansion involves an infinite sum over all pairs $\left(c,d\right)$ of coprime integers with $c> 0$, together with the pair $(c,d)=(0,1)$. The pair $\left(c,d\right)$ also determines the integers $a,b$ through the $\rm{SL}(2,\mathbb{Z})$ condition $ad-bc=1$. The expansion also involves a finite sum over the integers $\mu$ and $m$,  where $\mu=-\nnn+1,\ldots, \nnn$, while $m$ satisfies $0\leq 4\nnn m< \mu^2$ (the choice $\mu=0$, $m=0$ is not included in the sum).
Moreover, $\tilde c_{(\mu)}$ is a coefficient depending on the spectral-flow invariant combination $4\nnn m-\mu^2$ but independent of the other parameters or chemical potentials, which encodes the degeneracy of polar states (see below). Finally, $\Theta_{(\mu,\nnn )}$ is defined as  
\be
\Theta_{(\mu,\nnn )}\left(z,\tau\right) \,=\, \sum_{\eta\in\mathbb Z} \exp\left[{2\pi \ii \tau \,\frac{\left(\mu+ 2\nnn \eta\right)^2}{{4\nnn }} } + {2\pi \ii z\,(\mu + 2\nnn  \eta)}\right]\,.
\ee

 The Farey-tail expansion can be seen as a sum over saddle-point contributions having an interpretation as Euclidean geometries in type IIB string theory. To make this more manifest, we express~\eqref{eq:Fareytail_exp} as
\begin{equation}
\label{eq:rewritingFTexp}
    \chi_{\rm RR}\left(\tau,z\right) \,=\, -2\pi\ii\sum_{\left(c,d\right)}\,\sum_{\mu=-\nnn+1}^{\nnn }\,\,\sum_{0\leq m<\frac{\mu^2}{4\nnn } }\,\,\sum_{\eta\in\mathbb Z}\,\frac{{\tilde c}_{(\mu)}}{\left(c\tau+d\right)^3}\,\rme^{-I}\,,
\end{equation}
where 
\begin{equation}
\label{eq:actionfromFTE}
    I(\tau,z) \,=\, -2\pi \ii\left[ \left(m-\frac{\mu^2}{4\nnn} + \frac{1}{4\nnn }\left(\mu + 2\nnn  \eta \right)^2\right)\frac{a\tau +b}{c\tau +d} + \left(\mu + 2\nnn  \eta\right) \frac{z}{c\tau +d} - \nnn  \frac{cz^2}{c\tau + d} \right]\,
\end{equation}
is the action of the saddle identified by the quantum numbers $m,\mu,\eta,c,d$. When $\nnn$ is large, there are plenty of choices such that $I$ is $\mathcal{O}(\nnn)$. In the following we will focus on these cases, which are those that can possibly be matched by supergravity solutions. We will also assume that $\tilde c_{(\mu)}$ has a slower scaling with $\nnn$, so that it can be ignored at leading order. This has been verified for $m=0$~\cite{Dijkgraaf:2000fq}, and it is also true for the states to be considered in the next section.

Consider then the contribution of a saddle with action $I$.
Comparing the saddle-point approximation of \eqref{eq:ellipticgenus} with \eqref{eq:rewritingFTexp} yields the supersymmetric quantum statistical relation
\begin{equation}\label{eq:SUSYQSR}
 I\,=\,-2\pi \ii \left[\tau \left(L_0-\frac{\nnn}{4}\right)+ \left(z+\frac12\right) J_0 -\frac{1}{2}{\tilde J}_0\right]- {\cal S}\,.
\end{equation}

The expectation value of the left-moving charges associated with a given saddle can be evaluated by varying $I(\tau,z)$, \begin{equation}\label{eq:charges_Farey}
\begin{aligned}
L_0 - \frac{\nnn}{4}\,&=\, - \frac{1}{2\pi \ii}\frac{\partial I}{\partial \tau} \,=\, \frac{1}{(c\tau+d)^2}\left[m - \frac{\mu^2}{4\nnn  } + \frac{1}{4\nnn }\left( \mu + 2\nnn (\eta - cz)\right)^2 \right]\,,\\[1mm]
J_0 &\,=\, - \frac{1}{2\pi\ii}\frac{\partial I}{\partial z} \,=\,  \frac{\mu+ 2\nnn  (\eta - cz)}{c\tau+d}\,.
\end{aligned}
\end{equation}
These relations can be used to solve \eqref{eq:SUSYQSR} for the indicial entropy carried by the saddle, $\mathcal{S}+\pi \ii (J_0-\tilde J_0)$.
A basic observation is that the expressions \eqref{eq:charges_Farey} imply
\begin{equation}\label{eq:relation_charges}
4 {\nnn }\left(L_0-\frac{N}{4}\right)-J_0^2\,=\, \frac{4{\nnn } m-\mu^2}{(c\tau+d)^2}\,,
\end{equation}
independently of $\eta$ and $z$.
The combination on the left-hand side is the one appearing in the charged Cardy formula for two-dimensional CFT's. Recalling that $4{\nnn } m-\mu^2<0$ by assumption, we see that for $c\neq 0$, the regime $\tau\to -d/c$ from the upper half of the complex $\tau$ plane leads to an exponential $\mathcal{O}{(\rme^{\nnn})}$ growth of states. This indicates a black hole saddle. The case $c=0$ has no such exponential growth, and corresponds to saddles that are not sufficiently massive to form a black hole, i.e.\ that are below the black hole threshold. 

A central feature of the Farey tail expansion formula is that the full elliptic genus can be obtained by starting from the contribution of the  $c=0$ states and summing over their modular images obtained by acting with an ${\rm SL}(2,\mathbb{Z})$ element parameterized by $a,b,c,d$. This is encoded in the sum over $(c,d)$ with $c>0$.
In the following we extract some more information on the $c=0$ and $c\neq 0$ saddles, just based on their contribution to the elliptic genus.

\paragraph{Below the black hole threshold ($c=0$).} 
 Setting $(c,d)=(0,1)$ in \eqref{eq:actionfromFTE} and using that then $a= 1$ while $b$ drops out,\footnote{In this case $b$ contributes to $I$ as $2\pi\ii b$ times an integer, and the action is defined modulo shifts in $2\pi \ii \mathbb{Z}$.} we recover the contribution known as the {\it polar part} of the D1-D5 elliptic genus~\cite{Dijkgraaf:2000fq},
\begin{equation}
    \chi_{\rm  polar}\left(\tau,z\right) \,=\, - 2\pi\ii \sum_{\mu=-\nnn+1}^{\nnn }\,\,\sum_{0\leq m<\frac{\mu^2}{4\nnn } }\,\,\sum_{\eta\in\mathbb Z}\,\,{\tilde c}_{(\mu)}\,\rme^{-I}\,,
\end{equation}
with
\begin{equation}
\label{eq:actionpolarstates}
-\frac{I}{2\pi \ii} = \left(m-\frac{\mu^2}{4\nnn } + \frac{1}{4\nnn }\left(\mu + 2\nnn  \eta \right)^2\right)\tau + \left( \mu + 2\nnn \eta\right) z \,.
\end{equation}
This is the basic building block of the D1-D5 elliptic genus, since the rest can be reconstructed taking its modular images. The states contributing to $\chi_{\rm polar}$ are referred to as {\it polar states}.
 
Since the action is linear in the chemical potentials, the charges \eqref{eq:charges_Farey} coincide with the respective coefficients,
\begin{equation}\label{eq:charges_polarstates}
\begin{aligned}
L_0 - \frac{\nnn}{4}\,&=\,  m - \frac{\mu^2}{4\nnn  } + \frac{1}{4\nnn }\left( \mu + 2\nnn \eta\right)^2\,,\\[1mm]
J_0 &\,=\,  \mu+ 2\nnn  \eta\,.
\end{aligned}
\end{equation}
From these expressions, it is clear that $\eta$ parametrizes an integral spectral flow transformation in the left-moving sector, with the integers $(m, \mu)$ being the values of the charges $(L_0-\tfrac{\nnn}{4}, J_0)$ before spectral flow. The relation \eqref{eq:relation_charges} gives the inequality
\begin{equation}
4 {\nnn }\left(L_0-\frac{\nnn}{4}\right)-J_0^2\,=\, 4{\nnn } m-\mu^2 <0\,,   
\end{equation}
which as already noticed indicates that the polar states lie below the black hole threshold.

It is also clear that the indicial degeneracies at fixed left-moving charges are given by the $\tilde c_{(\mu)}$, since the indicial entropy vanishes at the classical $\mathcal{O}(\nnn)$ order, $\rme^{\mathcal{S}+\pi\ii(J_0-\tilde J_0)}= 1$.

The set of polar states that carry macroscopic $\mathcal{O}(\nnn)$ action and charges are expected to correspond to horizonless supergravity solutions. We will discuss  a family of such solutions in the next section.

\paragraph{Above the black hole threshold ($c\neq 0$).} Let us consider now the configurations with $c\neq 0$. In this case, we can use the identity
\begin{equation}\label{eq:rewriteSL2}
\frac{a\tau + b}{c\tau +d} \,=\, \frac{a}{c}- \frac{1}{c\left(c\tau +d\right)}
\end{equation}
to rewrite the expression for the saddle-point action $I$ in a convenient fashion,
\begin{equation}\label{eq:I_EG_BHs}
-\frac{I}{2\pi\ii}\,=\, \frac{a}{c}\left(m-\frac{\mu^2}{4\nnn}+ \frac{1}{4\nnn} \left(\mu+2\nnn\eta\right)^2\right)+\frac{1}{c\left(c\tau+d\right)}\left[\frac{\mu^2}{4\nnn} - m-\nnn\left(cz-\eta-\frac{\mu}{2\nnn}\right)^2\right]\, .
\end{equation}
At fixed values of the other variables, the saddles labeled by different values of $(c,d)$ compete between each other. As already noted, to each of them there is associated a Cardy-like limit $\tau \to -d/c$, such that the respective action is enhanced due to the pole in the second term. There is thus a rich pattern of phase transitions \cite{Dijkgraaf:2000fq}. The saddle with the fastest-growing action in the Cardy limit is obtained for $c=1$, $\mu = \nnn$, $m=0$.

 Note that one can recover the $c=0$ formulae of the previous paragraph starting from the $c\neq 0$ formulae above and formally taking the $c\to 0$ limit. In particular, this can be implemented in the saddle-point action~\eqref{eq:I_EG_BHs}. From \eqref{eq:rewriteSL2}, it is clear that the limit is finite although the two terms on the right hand side diverge if taken separately.

We can also consider the saddle-point contribution in a microcanonical ensemble where all charges apart for $\tilde J_0$ are fixed. 
Legendre transforming the above expression with respect to $\tau$ and $z$ gives the following function of the charges $L_0-\frac{\nnn}{4}$ and $J_0$,
\begin{equation}
\begin{aligned}\label{eq:S_Farey}
{\cal S}\,&=\,2\pi \,\frac{\sqrt{\mu^2-4\nnn m}}{c \nnn} \sqrt{\nnn \left(L_0-\frac{\nnn}{4}\right)-\frac{J_0^2}{4}}\\[1mm]
&\quad +2\pi \ii \left[\frac{{\tilde J}_0-J_0}{2}- \left(\eta+\frac{\mu}{2\nnn}\right)\frac{J_0}{c}+\frac{d}{c}\left(L_0-\frac{\nnn}{4}\right)+\frac{a}{c}\left(m-\frac{\mu^2}{4\nnn}+ \nnn \left(\eta+\frac{\mu}{2\nnn}\right)^2\right)\right]\,.
\end{aligned}
\end{equation}
This is a complex expression, depending on the choice of integers $\mu,m$ and $c,d$. If the charges are fixed to real values, the  second line is a purely imaginary contribution which generically is not an integer multiple of $2\pi \ii$, so it gives rise to a non-trivial phase when considering the indicial degeneracies $\rme^{\mathcal{S}+\pi\ii (J_0-\tilde J_0)}$.
Assuming real charges, the choice of integers $c,\mu,m$ that maximizes the real part of $\mathcal{S}$, namely the first line of~\eqref{eq:S_Farey}, is again $c=1$, $\mu = \nnn$, $m=0$. In this case, up to a trivial multiple of $2\pi \ii$,
\begin{equation}
{\cal S}\,=\,2\pi \sqrt{\nnn \left(L_0-\frac{\nnn}{4}\right)-\frac{J_0^2}{4}}
+\pi \ii \tilde J_0\,.
\end{equation}
This corresponds to the BTZ black hole saddle. In the next section, we are going to match a much larger family of gravitational saddles.

\section{Holographic match}
\label{sec:matchFareyTail}

We now interpret the  family of Euclidean AdS$_3\times S^3$ orbifolds presented in section~\ref{sec:decoupling2center} in the light of the Farey-tail expansion. We first discuss the dictionary between the gravitational and CFT variables, then we match the action of horizonless and black hole saddles in turn.

\subsection{The partition function}

It should be clear that one can identify the gravitational index of type IIB string theory on AdS$_3\times S^3\times \Kt$, whose saddles have been discussed in section~\ref{sec:uplift_and_decoupling}, and the D1-D5 elliptic genus \eqref{eq:ellipticgenus} reviewed in section~\ref{sec:CFT2}.
We summarize here the complete map relating the variables used in the two descriptions, although part of these relations have already appeared above.

Comparing the AdS$_3$ superalgebra relations \eqref{eq:3Dsuperalgebra} with the CFT superalgebra \eqref{eq:GG}, we identify the generators as
\begin{equation}
\label{eq:mapcharges}
 \mathcal{Q} = \tilde G_0^-\,,\quad\ \bar{\mathcal{Q}}=\tilde G_0^+\,,\quad\ J_- =\frac{J_0}{2}\,,\quad\ J_+ = -\frac{\tilde J_0}{2}\,,\quad\  N_{\rm P} = L_0 - \tilde L_0\,,\quad\ E = L_0+\tilde L_0 - \frac{\cc}{12}\,,
\end{equation}
where the gravitational quantities appear on the left hand side and the CFT generators on the right hand side. 
 The CFT central charge $\cc = 6 \nnn$ is given by 
\be
 \nnn \,=\, N_{\rm D1}N_{\rm D5} +1\,.
\ee

As already seen (recall \eqref{eq:tau}, \eqref{eq:omega_firsttime}, \eqref{eq:defz}), the chemical potentials are related by 
\begin{equation}
\label{eq:mappotentials}
    \varphi^3 = 2\pi \ii \tau\,,\ \qquad \beta \,=\, \pi \ii (\bar\tau-\tau) \,,\ \qquad \omega_+ = 4\pi \ii \tilde\omega\,,\qquad\ \omega_- =4\pi \ii\omega\, \quad\ \Rightarrow\quad\ \omega_2 = -2\pi\ii z\,.\quad
\end{equation}
We recall that the CFT constraint $\tilde \omega = \frac12$ given in \eqref{eq:omegatildeomega} provides the precise counterpart of the condition \eqref{eq:SUSYconstr} defining the gravitational index. 
The gravitational index $Z$ in the mixed ensemble of fixed $N_{\rm D1},N_{\rm D5}$ is then identified with the D1-D5 elliptic genus $\chi_{\rm RR}$ as
\begin{equation}
    Z\left(\omega_2 = -2\pi \ii z, \varphi^3 = 2\pi\ii\tau\right) \,=\, \chi_{\rm RR}\left(\tau,z\right)\,.
\end{equation}
Notice that the statement that the elliptic genus is independent of $\bar\tau$ (in a setup where the remaining variables are $\tau,z$) is equivalent to the fact that the gravitational index is independent of $\beta$ (in a setup where the remaining variables are $\varphi^3,\omega_2$).

\subsection{Horizonless saddles}

We can now demonstrate the match of the horizonless orbifold saddles of section~\ref{sec:decoupling2center} with a set of polar states in the Farey-tail expansion of the D1-D5 elliptic genus. Recall that these horizonless solutions are specified by two integers $k$ and $p$, besides the brane numbers $N_{\rm D1}$, $N_{\rm D5}$. The integers $k$ and $p$ specify the circle that collapses in the bulk as in \eqref{eq:U(1)_soliton}. Equivalently, they determine the orbifold action as in~\eqref{eq:orbifold_soliton}. Also, $k$ is identified with one of the three flat connections, cf.\ \eqref{eq:q3_is_k}, the other two being determined as in \eqref{eq:q1q2_soliton}.

Using the dictionary~\eqref{eq:mapcharges}, the charges of the two-center soliton \eqref{eq:charges_soliton2} read in CFT language
\be\label{eq:charges_GLMTstates}
L_0 - \frac{\nnn}{4}\,=\, N_{\rm P}\,=\, \frac{\nnn}{\korb}\frac{p(p-1)}{\korb} \,, \qquad J_0 \,=\, \frac{\nnn}{\korb} (2p-1)\,,\qquad \tilde L_0 - \frac{\nnn}{4}\, =\,0\,,\qquad \tilde J_0 \, =\, -\frac{\nnn}{\korb}\,,\quad
\ee
and satisfy the constraint
\begin{equation}
4 {\nnn }\left(L_0-\frac{\nnn}{4}\right)-J_0^2\,=\, -\tilde J_0^2 \,.
\end{equation}
Here we have taken into account the one-loop corrections and quantization conditions discussed at the end of section~\ref{sec:uplift_and_decoupling}, in particular we are assuming \eqref{eq:quantum_correction}, together with the divisibility condition
$
\frac{\nnn}{\korb}\in \mathbb{Z}
$
given in \eqref{eq:qcsol1},
and the one in \eqref{eq:qcsol3}.
 In turn, after decomposing $p$ as
 \be\label{eq:decomp_p}
 p \,=\, \eta \korb + p'\,,\qquad \text{with}\quad \eta \in \mathbb{Z}\,,\quad \   p'= - \left\lfloor \frac{k}{2} \right\rfloor+1 ,\ldots, \left\lfloor \frac{k+1}{2} \right\rfloor\,,
 \ee
we have that \eqref{eq:qcsol3} implies
 \be\label{eq:extra_condition}
 \frac{p'(p'-1)}{\korb}
 \,\in\, \mathbb{Z}\,.
 \ee
The integer $\eta$ is interpreted as an integral spectral flow parameter in the CFT, while $p'$ can be seen as parameterizing a fractional spectral flow~\cite{Giusto:2012yz}. 

Comparing the expressions for the left-moving charges in \eqref{eq:charges_GLMTstates} with~\eqref{eq:charges_polarstates}, we infer that these supergravity configurations can be interpreted as polar states with left-moving quantum numbers
\be\label{eq:mum_GLMT}
\mu \,=\, \frac{\nnn}{\korb}(2p'-1) \,, \qquad\qquad m \,=\, \frac{\nnn}{\korb} \frac{p'(p'-1)}{\korb} \,.
\ee
Note that we have chosen the range of $p'$ in \eqref{eq:decomp_p} so that $\mu\in [1-N,N]$. 
Plugging these values in \eqref{eq:actionpolarstates}, we obtain the action
\begin{equation}
I\,=\,-2\pi\ii\,\frac{ N}{k}\left[\frac{p(p-1)}{k}\,\tau+\left(2p-1\right)z\right]\,.
\end{equation}
which precisely matches the on-shell action \eqref{eq:solitonEG} of two-center solitons.
Note that this satisfies the shift relation $I_{k,\,p}(\tau,z+\tau) = I_{k,\,p+k}(\tau,z)+2\pi\ii N(\tau+2z)$, which follows from the spectral flow transformation $\eta \to\eta+1$ applied to \eqref{eq:actionpolarstates}.

We thus conclude that two-center solitons do contribute to the type IIB path integral computing the D1-D5 elliptic genus. As we have seen, the decoupling limit of these solutions yields  $({\rm AdS_3}\times S^3)/\mathbb{Z}_{\korb}$ geometries, generically with orbifold singularities. The CFT$_2$ interpretation of these states was discussed in~\cite{Giusto:2012yz}. They are obtained starting from a set of chiral primaries (namely, RR ground states) with equal ${\rm U}(1)$ R-charges $J_0 = \tilde J_0 = -\frac{\nnn}{\korb}$, where $\korb$ is a divisor of $\nnn$ by construction,\footnote{This comes from the fact that the chiral primaries are constructed by acting with a certain operator on the NS vacuum $\nnn/\korb$ times.}
 and acting in the left-sector with fractional spectral flow transformations having parameter $\frac{p}{\korb}=\eta + \frac{p'}{\korb}$. 
Taking $p$ to be a multiple of $k$, this family reduces to the states previously discussed in~\cite{ Giusto:2004id, Giusto:2004ip}, where only integer spectral flow transformations were considered.

\subsection{Black hole saddles}

We now consider the modular images of the polar states discussed above. We are going to show that they correspond to the general family of black hole orbifolds of section~\ref{sec:decoupling2center}.
In order to do so, we compare  the modular-transformed action~\eqref{eq:I_EG_BHs} with the black hole orbifold action~\eqref{eq:BHEG}, and verify that the latter exactly reproduces the former, provided we choose the discrete parameters characterizing the black hole orbifold so that they are suitably related to the integers describing the modular images of the polar states.

Plugging the values \eqref{eq:mum_GLMT} of $\mu,m$ in the action~\eqref{eq:I_EG_BHs} for the modular images of the polar states, we obtain
\begin{equation}\label{eq:I_modular_images}
I \,=\, -2\pi\ii\nnn \left[ \frac{a p(p-1)}{c\,\korb^2} + \frac{1-  \left(2\korb cz -2p + 1\right)^2}{4 \korb^2 c\left(c\tau+d\right)} \right]\, .
\end{equation}
This is matched by our black hole orbifold action \eqref{eq:BHEG} if we choose the quantities appearing there as
\be\label{eq:identif_nc_etc}
n 
\,=
\,c\,\korb   \,,\qquad\ q^3\,=\, -d\,\korb  \,,\qquad\ q^1 \,=\, -a\, N_{\rm D1} \,\frac{p(p-1)}{\korb} \,,\qquad\ q^2 \,=\, -a\,N_{\rm D5}\,\frac{p(p-1)}{\korb }\,.
\ee
 The match of the part of the action depending on the chemical potentials $\tau,z$ is immediate. We then focus on the constant phase term, namely the first term in \eqref{eq:I_modular_images},
\be\label{eq:phase_Farey}
-N\, \frac{a p(p-1)}{c\,\korb^2} 
\ee
(which is $-\frac{a}{c}$ times the momentum $N_{\rm P}$ of the soliton)
 and show that it is equal to the $\Psi$ term in~\eqref{eq:BHEG}, upon taking into account the correction~\eqref{eq:correction_Psi}. Since the match is non-trivial, we provide a detailed proof. 
 
Using \eqref{eq:identif_nc_etc} in the  expression \eqref{eq:phase_term} for $\Psi$,  and implementing~\eqref{eq:correction_Psi},  yields
\be
\begin{aligned}
\Psi 
\,&=\, -\frac{\nnn}{\korb}\,\frac{a p(p-1)}{c\korb } \left[2  + (w_0^1)^2 w_0^2 \,p(p-1)(n+p-1) ad\right].
\end{aligned}
\ee
This agrees with \eqref{eq:phase_Farey}  if 
\be\label{eq:to_be_simplified}
\frac{\nnn}{\korb}\,\frac{p(p-1)}{c\korb } \left[1  + (w_0^1)^2 w_0^2 \,p(p-1)(n+p-1) ad\right] \,\in\, \mathbb{Z}\,,
\ee
since this expression then yields a trivial phase which does not contribute to $\rme^{-I}$.
In order to see that~\eqref{eq:to_be_simplified} is true we need to use the relations satisfied by $w_0^1,w_0^2$, which were given in~\eqref{def:w}, \eqref{eq:quantization_w}. In particular, 
using that $w_0^1$ is a multiple of  $\frac{1}{{\rm gcd}(n,\,p-1)}$ while $w_0^2$ is a multiple of $\frac{1}{{\rm gcd}(n,\,p)}$, we see that the coefficient of  $ad$ in \eqref{eq:to_be_simplified}, where $(w_0^1)^2 w_0^2$ is multiplied by $p(p-1)(n+p-1)$, has the form of an integer divided by $c$. It follows that we can replace  $ad$ by 1, since $ad = 1 +bc$ and the $bc$ term gives rise to a trivial phase. 
In order to conclude, recalling that $\frac{N}{\korb}$ is integer, it is sufficient to prove that
\be
\frac{p(p-1)}{n} \left[1  + (w_0^1)^2 w_0^2 \,p(p-1)(n+p-1)\right] \,\in\, \mathbb{Z}\,,
\ee
where we used $c\korb=n$.
Using~\eqref{def:w}, the second factor can be expressed as:
\be\label{eq:integer_expression}
\begin{aligned}
1+ (w_0^1)^2 w_0^2 \,p(p-1)(n+p-1)  \,&=\, 1-(1-nw_2^2)(1-n(w_0^1+w_1^1))(1-n w_1^1)
 \\[1mm]
 \,&=\, n  (w_0^1u_0+w_1^1 u_1 + w_2^2 u_2) 
 \,,
\end{aligned}
\ee
where we may take $u_0=1-n w_2^2$, $u_1=2-n w_0^1-n w_2^2$, $u_2=1-n w_1^1+n^2 w_0^1 w_1^1$. These expressions for $u_0,u_1,u_2$ are not important, we just need to note that they are integers since $w_0  ^1,w_1^1,w_2^2$ are multiples  of $\frac{1}{{\rm gcd}(n,p)}$ or $\frac{1}{{\rm gcd}(n,p-1)}$ (again, recall~\eqref{eq:quantization_w}) and each of them is multiplied by $n$. Similarly, the integer expression in the second line of~\eqref{eq:integer_expression} has the property of remaining integer also after it is multiplied by $\frac{p(p-1)}{n}$. 
This concludes our proof.

\medskip 

We have thus matched an infinite family of supersymmetric orbifolds of Euclidean BTZ $\times S^3$ with saddles arising in the Farey-tail expansion of the D1-D5 elliptic genus above the black hole threshold. As discussed in section~\ref{sec:decoupling2center}, the solutions are characterized by the integers $n,p,q^3$. These specify the Killing vector generating the circle that collapses in the bulk as in~\eqref{eq:xi_orbifold_general}, and the orbifold action as in \eqref{eq:BTZxS3orbifolds}. As we have discussed, depending on the possible common factors of these integers, the orbifold may be freely acting or singular. The orbifold singularities, when present, appear at either one (or both) of the poles of the horizon~$S^3$. 

The identification~\eqref{eq:identif_nc_etc} provides the most convenient parameterization of $n,q^3$: this is in terms of a possible common factor $k$  and coprime integers $(c,d)$, with $c>0$ (the choice $c=0$, $d=1$ giving back the horizonless saddles above).  
For $k=1$, we recover the $(c,d)$ family of black hole saddles discussed in~\cite{Dijkgraaf:2000fq}---see also \cite{Kraus:2006nb}---which in addition depend on the shifts by $p\in\mathbb{Z}$ parameterizing the possible twistings of $S^3$ 
 that are compatible with supersymmetry. Here we have derived this family of saddles, along with their  saddle-point action, taking the decoupling limit of supersymmetric non-extremal solutions asymptotic to $S^1\times S^1\times \mathbb{R}^4$. If we further set $c=1,d=0$, $p=0$ we recover ${\rm BTZ}\times S^3$, with on-shell action
\be
I_{k=c=1,\,d=p=0} \,=\, - \frac{\pi \ii N}{2}\frac{1-(2z+1)^2}{\tau}\,=\, - \frac{\pi \ii N}{2}\frac{1-4\omega^2}{\tau}\,,
\ee 
where we have used the second in \eqref{eq:omegatildeomega} in order to express it in terms of $\omega$ instead of $z$.

The $k>1$ saddles enhance the  explicitly known saddles of the D1-D5 elliptic genus above the black hole threshold.
They
can be seen as ${\rm SL}(2,\mathbb{Z})$ modular images of the horizonless solutions of~\cite{ Giusto:2012yz}, which correspond to $({\rm AdS}_3 \times S^3)/\mathbb{Z}_k$ orbifolds, rather than just ${\rm AdS}_3 \times S^3$. In particular, taking $c=1$, $d=0$, $p=0$,  the orbifold acts on the  time circle and leaves the spatial circle in Euclidean AdS$_3$ untouched; this is the same action as in the horizonless solutions, with the space and time circles reversed.
The action reads in this case 
\begin{equation}
I_{c=1,\,d=p=0} \,=\, -\frac{\pi\ii N}{2}\,   \frac{1-  \left(2\korb z + 1\right)^2}{\korb^2 \tau} \, .
\end{equation}


\section{Comments on black ring saddles}\label{sec:black_ring}

In the previous sections, we considered a family of two-center solutions of five-dimensional supergravity uplifted to Type IIB on $S^1\times {\rm K}3$, and discussed how they contribute after a suitable decoupling limit to the elliptic genus of the dual D1-D5 CFT, thereby reproducing a large number of terms of its Farey-tail expansion. One may wonder whether the terms in this expansion that remain to be matched could be associated with configurations with more than two centers. In this section, we provide evidence that this is not the case, at least within the setup considered here. To this end, we analyse a relevant subclass of three-center solutions and identify an obstruction that we expect to persist generally for multi-center configurations with three or more centers.

Three-charge saddles of the five-dimensional gravitational index with non-spherical horizon topology can be constructed within the general multi-center ansatz of section~\ref{sec:bubbling_saddles}. An interesting subclass consists of solutions with $S^1\times S^2$ horizon, which provide finite-temperature, yet supersymmetric, deformations of the well-known BPS black rings of~\cite{Elvang:2004rt,Elvang:2004ds,Emparan:2006mm}. As shown in appendix~\ref{app:lens_ring_5D}, these arise as three-center configurations. Their on-shell action is obtained from \eqref{eq:3centergeneral} by setting $p_1=1$.\footnote{More precisely, the action for the finite-temperature deformation of the BPS black rings of~\cite{Elvang:2004rt,Elvang:2004ds,Emparan:2006mm} is obtained from \eqref{eq:3centergeneral} by setting $p_1=1$, $p=0=\tilde q^I$, and $n=1$, with $\tilde q_1^I$ playing the role of the ring dipole charges. Here, then, we consider a slightly more general solution depending also on such integers.}
Using the uplift formulae \eqref{eq:upliftformulae}, these can be embedded in type IIB supergravity on $S^1\times {\rm K}3$. Taking the decoupling limit described in section~\ref{sec:decoupling}, then, yields a family of three-center configurations asymptotic to AdS$_3\times S^3\times {\rm K}3$, satisfying the boundary conditions necessary for contributing to the elliptic genus \eqref{eq:ellipticgenus}.\footnote{A decoupling limit for supersymmetric multi-center saddles with non-trivial horizon topologies was recently discussed in~\cite{Boruch:2025sie} (building on~\cite{deBoer:2008fk}) in a different setup, where it leads to asymptotically $\mathrm{AdS}_3\times S^2$ geometries contributing to the elliptic genus of a dual $(0,4)$ CFT.} The corresponding saddle-point action follows from \eqref{eq:3centerAdS3}:
\begin{equation}
\label{eq:BRAdS3}
    I \,=\, \frac{2\pi \ii N_{\rm D1}N_{\rm D5}}{n\,q_1^3}\left(1+\check z\right) - \frac{2\pi \ii}{n \,q_1^3}\Biggl[\left(q_1^1 N_{\rm D5}+ q_1^2 N_{\rm D1}\right) -q_1^1q_1^2\frac{\check \tau}{1+\check z}\Biggr] \left(\check \tau - q_1^3\check z\right)
     + 2\pi\ii \Psi\,,
\end{equation}
where
\begin{equation}
    \check\tau \,=\, n\tau - q^3\,,\qquad\qquad \check z \,=\, n z -p\,,
\end{equation}
and
\be
\begin{aligned}
\Psi \,&=\, \frac1n\Biggl[ q^1 N_{\rm D5} + q^2 N_{\rm D1} 
\\[1mm]
&\qquad -w_0^2\left[ q_1^1 q^2 q^3 + q_1^2 q^1 q^3 + q_1^3 q^1 q^2 -p \left(q^1 q_1^2 q_1^3 + q^2 q_1^1 q_1^3 + q^3 q_1^1 q_1^2\right) + p^2 q_1^1 q_1^2 q_1^3\right] \Biggr]\,,
\end{aligned}
\ee
with the integer $w_0^2$ being defined through eq.~\eqref{eq:w233center}.
Let us comment on the structure of the action above. The main obstruction to interpreting \eqref{eq:BRAdS3} as a saddle-point of the elliptic genus is the presence of a pole in the chemical potential $z$, located at $z\to (p-1)/n$.\footnote{We are also assuming that $q_1^1,q_1^2,q_1^3$ are non-vanishing, and that $q^1$ and $q_1^1$ grow as $N_{\rm D1}$, while $q^2,q_1^2$ grow as $N_{\rm D5}$, as for the two-center case, so that none of the terms in \eqref{eq:BRAdS3} is subleading in the expansion at large $N_{\rm D1} N_{\rm D5}$.} By contrast, each term in the Farey-tail expansion of the elliptic genus has the form \eqref{eq:actionfromFTE}, meaning that it can develop poles only at $\tau\to -d/c$. This follows from modularity: the contributions of states above the black hole threshold are reconstructed as modular images of the polar sector, whose action is linear in both $\tau$ and $z$. The only denominator that can be generated by a modular transformation is $(c\tau+d)$. Therefore, more generally, an action containing $z$ in the denominator cannot reproduce any term in the Farey-tail expansion.\footnote{This obstruction also arises for the black lens configurations considered in appendix~\ref{app:lens_ring_5D}, where $p_1=-1$. Indeed, the more general action \eqref{eq:3centerAdS3} contains a denominator $(1-p_1)\check\tau+q_1^3\left(1+\check z\right)$, which for $p_1=-1$ depends on both $\tau$ and $z$ and is not of the form $c\tau+d$ characteristic of the Farey-tail saddle-point actions.}

\paragraph{Horizonless limit.} We next investigate whether multi-center configurations may contribute to the polar sector, i.e.\ below the black hole threshold. 
As we have explained, horizonless  configurations arise in the $n\to0$ limit of our general saddles. In appendix~\ref{app:lens_ring_5D}, we show that taking this limit starting from the asymptotically flat five-dimensional black ring geometries with arbitrary $n$, yields the Euclidean counterpart of a three-center (i.e.\ with two compact two-cycles) horizonless geometry (see e.g.~\cite{Bena:2007kg}). After the uplift and decoupling limit, this construction then produces an asymptotically AdS$_3\times S^3\times {\rm K}3$ soliton with two compact cycles. The action \eqref{eq:3centerAdS3} remains finite as $n\to0$, becoming 
\begin{equation}
\label{eq:BRsolitonAdS3}
\begin{aligned}
    I \,&=\, 2\pi \ii \Biggl[\Biggl(
    \frac{N_{\rm D1}N_{\rm D5}}{q_1^3}+ q_1^1 N_{\rm D5} + q_1^2 N_{\rm D1} - \frac{q_1^1 q_1^2 q^3}{q_1^3}\frac{q^3-q_1^3}{\left(1-p\right)^2}
    \Biggr)\,z 
    \\[1mm]
   \, &\,\qquad\, -\Biggl(  q_1^1 N_{\rm D5} + q_1^2 N_{\rm D1} + q_1^1q_1^2\frac{\left(2q^3 - p q_1^3\right)}{\left(1-p\right)}\Biggr)\,\frac{\tau}{q_1^3}\Biggr]\,.
    \end{aligned}
\end{equation}
This is linear in the chemical potentials $\tau$ and $z$, as it should in horizonless solutions. 
Hence the charges are directly read from the respective coefficients, $L_0 - \frac{N}{4} = -\frac{1}{2\pi\ii}\frac{\partial I}{\partial\tau}$ and $J_0 = -\frac{1}{2\pi\ii}\frac{\partial I}{\partial z}$. Linearity in the chemical potentials also characterizes the action of polar states contributing to the elliptic genus, recall \eqref{eq:actionpolarstates}.
Beyond that, a necessary condition for contributing to the polar sector is that the inequality $4\nnn (L_0-\frac{N}{4})- J_0^2 <0$ be satisfied. 
 It is possible to identify ranges of the parameters for which this inequality holds, thereby constraining the ring dipole charges $q_1^I$. 
Nonetheless, the argument of~\cite{Bossard:2019ajg} shows that horizonless geometries with more than one compact cycle, once uplifted to type IIB supergravity compactified on ${\rm K}3$, do not exist as supersymmetric configurations at a generic point of the moduli space. Therefore, they are not expected to provide protected states contributing to an index.\footnote{The analysis relies on the fact that the $1/4$-BPS solutions under consideration, belonging to theories with at least sixteen supercharges, also lie within a truncation with eight supercharges. A related argument was also developed in~\cite{Dabholkar:2009dq} for four-dimensional ${\cal N}=4$ supergravity. There, two-center configurations containing at least one (extremal) macroscopic horizon were shown to preserve supersymmetry only on submanifolds of the moduli space of codimension greater than one and, therefore, should not contribute to protected indices. Although the analysis was carried out in four dimensions, the authors argued that it should extend to five-dimensional supergravity. It would be interesting to revisit this argument to establish rigorously that the supersymmetric non-extremal black ring configurations considered in this section cannot contribute to the D1-D5 elliptic genus.} 



\section{Conclusions}\label{sec: conclusions}

In this paper, we have connected candidate saddles of the five-dimensional gravitational index with $S^1\times\mathbb{R}^4$ boundary conditions to a microscopic description. In order to do so, we have uplifted the five-dimensional supergravity theory to type IIB supergravity on $S^1\times M_4$, so that the five-dimensional black holes become six-dimensional black strings carrying D1-D5-P charges. Then we have taken a decoupling limit where the $S^1\times \mathbb{R}^4\times S^1\times M_4$ asymptotics are replaced by Euclidean  ${\rm AdS}_3\times S^3\times M_4$ asymptotics, where $S^3$ is twisted over the torus in AdS$_3$. The resulting solutions are candidate saddles for a gravitational index with AdS$_3$ boundary conditions, which can be studied using a microscopic CFT$_2$ description. We emphasize that both before and after the limit, the candidate saddles are complexified supersymmetric non-extremal configurations, where the Euclidean time circle maintains a finite length~\cite{Cabo-Bizet:2018ehj}. The extremal configurations are only reached if one takes a further limit where this circle is decompactified. 

Focusing on $M_4=\Kt$, we then compared the asymptotically AdS$_3$ candidate saddles with the  elliptic genus of the dual CFT$_2$, expressed as a sum over saddle-point contributions arising from the Farey-tail expansion. The comparison puts the original asymptotically flat saddles on more solid grounds. In particular, it informs us on which candidate gravitational saddles do indeed contribute to the D1-D5 elliptic genus, and which ones do not. Clearly, it would be desirable to also reach a purely gravitational understanding of the quantization conditions and further restrictions to be imposed so that candidate saddles actually contribute.

In this D1-D5 setup, we have established a precise match of an infinite family of saddles, originating from two-center configurations and characterized by three integers, comprising saddles both below and above the black hole threshold. 
Below the black hole threshold, we identified a family of polar states, corresponding to the Euclidean version of the horizonless solutions discussed in \cite{Giusto:2012yz}, which  in the dual CFT correspond to states obtained via fractional spectral flow. After the decoupling limit, these are $({\rm AdS}_3\times S^3)/\mathbb{Z}_k$ orbifolds.  Above the black hole threshold, we discussed the  SL$(2,{\mathbb Z})$ families of  black hole saddles that are obtained by taking the modular images of each of the polar states above. This generalizes the ${\rm SL}(2,\mathbb{Z})$ family of BTZ black holes discussed in~\cite{Maldacena:1998bw,Dijkgraaf:2000fq} by the orbifold parameter $k$, with the case of~\cite{Maldacena:1998bw,Dijkgraaf:2000fq} corresponding to $k=1$. For all these gravitational saddles, we provided the on-shell action in the relevant ensemble of fixed D1-D5 charges and specified how the orbifold preserves supersymmetry.

While we have derived the on-shell action from the one of the asymptotically $S^1\times \mathbb{R}^4$ configurations, it would certainly be possible to reproduce our expressions \eqref{eq:BHEG}, \eqref{eq:solitonEG} via a calculation performed directly in the asymptotically ${\rm AdS}_3 \times S^3$ setup. One way to do so would be to apply equivariant localization~\cite{BenettiGenolini:2023kxp}, for instance using the ten-dimensional approach of \cite{Couzens:2026xmi}, or starting from three-dimensional Chern-Simons $\mathsf{A}\wedge\diff \mathsf{A}$ supergravity and using equivariant localization of the corresponding $\diff\mathsf{A}\wedge\diff \mathsf{A}$ anomaly polynomial, adapting the method of~\cite{Cassani:2026teb,BenettiGenolini:2026cdw}. Relatedly, we observe that these expressions for the on-shell action do not receive higher-derivative corrections~\cite{Kraus:2006nb}.

Although the polar state saddles identified by our construction constitute a large family, arguably comprising the most general polar state saddles admitting a five-dimensional supergravity description, it certainly does not exhaust the full set of polar states predicted by the D1-D5 elliptic genus that may be reproduced in type IIB supergravity. It would  be interesting to identify these more general gravitational contributions.
In particular, it would be worthwhile to investigate how superstrata~\cite{Bena:2015bea}, as well as their orbifolded versions~\cite{Shigemori:2022gxf}, can contribute as polar states. 
This would require suitable restrictions on the charges (so that the states lie below the black hole threshold) and on the wave numbers (so that the solutions are globally well-defined after the Euclidean time is compactified, and a discrete family is identified out of configurations that a priori depend on arbitrary functions).
These would provide examples of saddles of the elliptic genus that admit a six-dimensional supergravity description but not a five-dimensional one. If this is indeed the case, modularity of the elliptic genus would imply the existence of an entire family of new ``black superstratum'' saddles. One might expect these to be constructed by exchanging the roles of the spatial and thermal circles in the Euclideanized version of the known superstrata, and subsequently taking their $(c,d)$ modular images with $c>1$, analogously to the construction presented in this paper for the $({\rm AdS}_3\times S^3)/\mathbb{Z}_k$ orbifolds. Such construction would lead to solutions carrying excitations around the thermal circle. 

It would furthermore be interesting to develop a worldsheet description of our index saddle configurations. A useful starting point may be offered by the recent construction of~\cite{Massai:2025nci},  suitably adapted to accommodate Euclidean backgrounds.

We also provided new evidence of some no-go results indicating that supersymmetric configurations with more than two centers do not contribute to the D1-D5 elliptic genus: we find  that the explicit form of the on-shell action of these saddles does not match any of the saddles predicted by the Farey-tail expansion. This includes both black ring and black lens saddles. However, one should recall that  the D1-D5 system is a very constrained highly supersymmetric setup, where the CFT$_2$ preserves $(4,4)$ supersymmetry. It would be interesting to consider other setups with different microscopic realizations, such as  elliptic genera of $(0,4)$ CFTs (see~\cite{deBoer:2006vg,Denef:2007vg} for discussions of possible Farey-tail expansions), and clarify under which conditions black ring and black lens saddles contribute.

Finally, we observe that some of the lessons learned in the present ${\rm AdS}_3/{\rm CFT}_2$ context  may help clarifying the structure of the gravitational path integral in higher dimensions.
In particular, there is a similarity between the discussion here and the one for ${\rm AdS}_5\times S^5$ black hole saddles~\cite{Cabo-Bizet:2018ehj,Choi:2018hmj,Benini:2018ywd}.  Although the modular symmetry that constrains the ${\rm AdS}_3/{\rm CFT}_2$ partition function is not present in the ${\rm AdS}_5/{\rm CFT}_4$ case, shifted and orbifolded black hole saddles have been shown to contribute in that case too~\cite{Aharony:2021zkr}, and matched with a dual field theory analysis~\cite{Benini:2018ywd,Cabo-Bizet:2019eaf,Aharony:2021zkr}. An important aspect of the ${\rm AdS}_3/{\rm CFT}_2$ story is the presence of orbifold singularities, implying string theory twisted sectors. The ${\rm AdS}_5$ orbifolds also allow for singular loci~\cite{Aharony:2021zkr}, and it would be interesting to investigate the implications these may have in the partition function. 
Another insight offered by the asymptotically ${\rm AdS}_3\times S^3$ saddles is that the family of orbifold black holes is connected with a horizonless solution via a limit where the orbifold parameter is sent to zero. The quantization conditions imposed in the horizonless solutions are then inherited in a suitably adapted form by the orbifold saddles. It would be interesting to investigate the higher-dimensional counterpart of this relation.


\subsection*{Acknowledgments}

We would like to thank Stefano Giusto, Ji Hoon Lee, Stefano Massai, Shanmugapriya Prakasam, Mart\'i Rossell\'o and Amitabh Virmani for interesting discussions. We also thank the Galileo Galilei Institute for Theoretical Physics, along with the participants, for the stimulating environment at the ``Pathways to Quantum Black Holes'' workshop, where part of this work was done. 
AR is supported by the University of Padua and the Fondazione Cariparo under the STARS@UNIPD 2025 programme (GRASBH -- The gravitational path integral, supersymmetric black holes and higher-derivative corrections). ET is supported by the National Research Foundation of Korea under the grant RS-2025-00518906.  ET would like to thank the Theoretical Physics group at the Korean Institute for Advanced Study (KIAS) in Seoul for hospitality during the initial stage of this project.

\appendix

\section{Ten-dimensional regularity analysis}
\label{app:10Dregularity}

The goal of this appendix is just to show that the regularity conditions \eqref{eq:regularitycond5D} and \eqref{eq:regularitycond5D2} can be derived from a ten-dimensional regularity analysis as well, serving this as cross-check of the analysis in \cite{Cassani:2025iix}. As already mentioned in section~\ref{sec:uplift_and_decoupling}, the main difference in the analysis is the geometrization of the U(1) gauge group associated to the Kaluza-Klein vector, $A^3$. As a consequence, the gauge transformation induced when going around the thermal circle is realized by a shift in the coordinate $y$ parametrizing the $S^1$, leading to the periodic identifications in \eqref{eq:periodicidentifications}, which we repeat here
\begin{equation}
\begin{aligned}
\left(t_{\rm E}, y, \psi, \phi\right)\sim& \left(t_{\rm E} +\beta, y-\ii \varphi^3-\ii\beta,\psi+\ii\omega_-, \phi+2\pi\right)\sim \left(t_{\rm E}, y+2\pi R_y, \psi, \phi\right) \\[1mm]
\sim& \left(t_{\rm E}, y, \psi +4\pi, \phi\right)\sim \left(t_{\rm E}, y, \psi+2\pi, \phi+2\pi\right)\,.
\end{aligned}
\end{equation}
A basis of $2\pi$-periodic Killing vectors generating evolution along each of the circles is
\begin{equation}\label{eq:basisvectors2}
\frac{\beta}{2\pi}\partial_{t_{\rm E}}+\frac{\varphi^3+\beta}{2\pi\ii}\partial_{y}+\frac{\ii\omega_-}{2\pi}\partial_{\psi}+\partial_{\phi} \,, \hspace{1cm} R_y \partial_y\,, \hspace{1cm} \partial_{\phi}+\partial_{\psi}\hspace{1cm} \partial_{\phi}-\partial_{\psi}   \,. 
\end{equation}
As it turns out, the following linear combinations of the basis vectors \eqref{eq:basisvectors2} degenerate along the $z$-axis (concretely, at the interval ${\cal I}_a$ joining the centers at $z_a$ and $z_{a+1}$):
\begin{equation}\label{eq:rodvectors2}
\xi_{{\cal I}_a}\,=\, \partial_{\phi}-\ii{\breve \omega}_{{\cal I}_a}\partial_{t_{\rm E}}-\chi_{{\cal I}_a}\partial_{\psi}+\left({\breve A}^3_{{\cal I}_a}-{\breve \omega}_{{\cal I}_a}\right)\partial_y\,,    
\end{equation}
where ${\breve \omega}_{{\cal I}_a}, \chi_{{\cal I}_a}$ and ${\breve A}^3_{{\cal I}_a}$ stand for the $\phi$-component of the associated one-forms evaluated at the rod ${\cal I}_a$, namely
\begin{equation}
{\breve \omega}_{{\cal I}_a}\,=\, -2\sum_{b\le a} w_b\,, \hspace{1cm} \chi_{{\cal I}_a}\,=\,1-2\sum_{b\le a}h_b\,, \hspace{1cm}  {\breve A}^3_{{\cal I}_a}\,=\, -2\ii\sum_{b\le a}\kk^3_b\, .
\end{equation}
Regularity imposes that the Killing vectors \eqref{eq:rodvectors2} have closed orbits, which is equivalent to demanding that \eqref{eq:rodvectors2} are given by a linear combination of the basis vectors \eqref{eq:basisvectors2} with integer coefficients. A suitable parametrization of these integers is 
\begin{equation}
\label{eq:rodvector6D}
\begin{aligned}
\xi_{{\cal I}_a}\,&=\, n_a \left(\frac{\beta}{2\pi}\partial_{t_{\rm E}}+\frac{\varphi^3+\beta}{2\pi\ii}\partial_{y}+\frac{\ii\omega_-}{2\pi}\partial_{\psi}+\partial_{\phi}\right) - q^3_a R_y \partial_y + p_a \left(\partial_{\phi}+\partial_{\psi}\right) \\[1mm]
\,&\,+ \left(1-n_a-p_a\right)\left(\partial_{\phi}-\partial_{\psi}\right)\,,
\end{aligned}
\end{equation}
which leads to the following regularity conditions
\begin{equation}\label{eq:regularitycond}
\begin{aligned}
\sum_{b\le a}\ii w_b\,&=\,\frac{n_a \beta}{4\pi}\,, \\[1mm]
\sum_{b\le a}h_b\,&=\,\frac{n_a}{2}\left(\frac{\ii\omega_-}{2\pi}+1\right)+p_a\,,\\[1mm] 
\sum_{b\le a}\kk^3_b\,&=\,\frac{n_a\,\varphi_3}{4\pi}+\frac{q^3_a R_y}{2\ii}\, ,
\end{aligned}
\end{equation}
precisely reproducing the regularity conditions found in \eqref{eq:regularitycond5D} and \eqref{eq:regularitycond5D2}, after identifying the fundamental charge of the KK vector as 
$
e^3\,=\,R_y^{-1}.
$

\section{Three-center saddles, black rings and black lenses}
\label{app:lens_ring_5D}

In this section we consider a family of three-center solutions of five-dimensional supergravity,  slightly generalizing the presentation in section 5 of~\cite{Cassani:2025iix}. These geometries possess two compact rods, ${\cal I}_1 = [z_1,z_2]$ and ${\cal I}_2 = [z_2,z_3]$, and we focus on the case in which the latter is a Euclidean horizon; the former, then, must be a non-trivial bubble.
The Killing vectors degenerating at such loci are given by
\begin{equation}
    \xi_{{\cal I}_1} = \left(1-p_1\right)\partial_{\phi_1} + p_1\,\partial_{\phi_2}\,,\qquad \xi_{{\cal I}_2} = n\,\partial_{\phi_0} + \left(1-n-p\right)\partial_{\phi_1} + p\,\partial_{\phi_2}\,,
\end{equation}
while the two semi-infinite rods ${\cal I}_{0,3}$ are as usual associated to $\xi_{{\cal I}_0} = \partial_{\phi_1}$ and $\xi_{{\cal I}_3} = \partial_{\phi_2}$. 
The center $z_1$ exhibits a conical singularity controlled by the integer $p_1$, with the choice $|p_1|=1$ leading to a regular point. This condition can be equivalently formulated in terms of the existence of a suitable vector $w^1 \equiv w_0^1 \partial_{\phi_0} + w_1^1 \partial_{\phi_1} + w_2^1\partial_{\phi_2}$, with $w^1_{0,1,2}\in\mathbb Z$, such that
\begin{equation}
\label{eq:3centrew1}
    {\rm det}\left(\xi_{{\cal I}_0} ,\xi_{{\cal I}_1},w^1\right) =1\,,
\end{equation}
where $\xi_{{\cal I}_0}$ and $\xi_{{\cal I}_1}$ are the two vectors whose orbits degenerate at $z_1$. Indeed, since 
$${\rm det}\left(\xi_{{\cal I}_0} ,\xi_{{\cal I}_1},w^1\right) = w^1_0 p_1\,,$$ then \eqref{eq:3centrew1} admits an integer solution for $w_0^1$ only if $p_1 = \pm 1$. The analysis of~\cite{Cassani:2025iix} shows that smoothness at the remaining two centers requires 
\begin{equation}
\label{eq:gcd3center}
    {\rm gcd}\left(n,p_1-p\right) =1 \,,\qquad {\rm gcd}\left(n,p-1\right)=1\,.
\end{equation}
If either condition fails, the corresponding center carries an orbifold singularity -- of order $\mathbb Z_{|{\rm gcd}\left(n,p_1-p\right)|}$ or $\mathbb Z_{|{\rm gcd}\left(n,1-p\right)|}$. As above, following~\cite{Colombo:2025yqy}, these smoothness conditions can be expressed as the existence of vectors $w^a\equiv w^a_0 \partial_{\phi_0} + w^a_1 \partial_{\phi_1} + w^a_2 \partial_{\phi_2}$, with $w^a_{0,1,2}\in \mathbb Z$ and $a=2,3$, satisfying
\begin{equation}
    {\rm det}\left(\xi_{{\cal I}_{a-1}} ,\xi_{{\cal I}_a},w^a\right) =1\,,\qquad \text{with}\,\,\,a=2,3\,.
\end{equation}
These determinant conditions reduce to
\begin{equation}
\label{eq:w233center}
    1= w_1^2 np_1 - w_0^2 \left(p_1\left(1-n\right) -p\right)\,,\qquad -1 = w_1^3 n - w_0^3\left(1-n-p\right)\,.
\end{equation}
By Bezout's lemma, integer solutions $w_{0,1}^{2}, w_{0,1}^{3}\in\mathbb Z$ exist when \eqref{eq:gcd3center} holds.\footnote{Notice that the three vectors $w^{1,2,3}$ are defined up to shifts $w^a \to w^a + n_1^a \xi_{{\cal I}_{a-1}} + n_2^a \xi_{{\cal I}_a}$, with $n_{1,2}^a \in \mathbb Z$. We have used this freedom to set $w^1_{1,2} =0= w^2_2 = w^3_2$. Then, we are left with $w^1_0 = p_1$, while the independent components of $w^{2,3}$ are determined by \eqref{eq:w233center}.} Although geometries with such orbifold singularities may still contribute to a gravitational index and could in principle be included in our analysis, in the remainder of this section we restrict, for simplicity, to configurations satisfying the smoothness conditions in \eqref{eq:gcd3center}, together with $p_1=\pm1$. This is, indeed, sufficient for the discussion in section~\ref{sec:black_ring}. Under these assumptions, the three-dimensional fixed locus of the Killing vector ${\xi}_{{\cal I}_1}$ is the lens space $L\left(|n|,p-p_1\right)$. It is not hard to show this, following~\cite{Cassani:2025iix,Colombo:2025yqy}. Since \eqref{eq:3centrew1} holds, the three vectors $\left(\xi_{{\cal I}_0},\xi_{{\cal I}_1},w^1\right)$ form a suitable basis, hence we can express $\xi_{{\cal I}_2}$ as the following linear combination, 
\begin{equation}
    \xi_{{\cal I}_2} = \mathtt q\,\xi_{{\cal I}_0} + p_1p\,\xi_{{\cal I}_1} + \mathtt p \,w^1\,,\qquad \mathtt q = 1-n-p_1p\,,\quad \mathtt p = n p_1\,.
\end{equation}
On the bubble fixed locus we have $\xi_{{\cal I}_1} =0$, while the two circles collapsing at its endpoints $z_{1,2}$ are generated by the vectors $\xi_{{\cal I}_{0,2}}$. As explained e.g. in~\cite{Cassani:2025iix,Colombo:2025yqy}, this describes the lens space $L\left(|\mathtt p|,\mathtt q\right)$. Then, for $p_1 = \pm 1$ one finds $L\left(|np_1|,1-n-p_1p\right)\cong L(|n|,p-p_1)$ (up to orientation). The same argument determines the topology of the Euclidean horizon, namely the three-dimensional fixed locus of the vector $\xi_{{\cal I}_2}$. In this case one finds that $\xi_{{\cal I}_3}$ can be expressed as the following linear combination of $\xi_{{\cal I}_{1,2}}$ and $w^2$:
\begin{equation}
\xi_{{\cal I}_3} = \left(p_1 + p\left(1-p_1\right)w_0^2\right) \,\xi_{{\cal I}_1} + \left(1-p_1\right)w_0^2\,\xi_{{\cal I}_2} + n\left(p_1-1\right) w^2\,.
\end{equation}
This allows us to conclude that the Euclidean horizon also describes a lens space, i.e. 
\begin{equation}
    L\left( |n\left(p_1-1\right)|,p_1 + p\left(1-p_1\right)w_0^2\right)\,,
\end{equation}
with $w_0^2$ determined by \eqref{eq:w233center}.
In the following we distinguish two relevant cases: black rings and black lenses.
\begin{itemize}
\item {\bf Black rings.} For
$p_1 =1$,
the Euclidean horizon reduces to $L(0,1) \cong S^1\times S^2$. When $n=1$ and $p=0$ this describes a complex non-extremal deformation of the BPS black ring of~\cite{Elvang:2004rt,Elvang:2004ds,Emparan:2006mm}. For $\big|n\big|>1$ and $p\in \mathbb Z$ subject to \eqref{eq:gcd3center}, the geometries describe an infinite family of non-singular orbifold solutions with the same horizon topology. 
\item {\bf Black lenses.} For 
$p_1 = -1 \,,$ $n=\pm 1\,,$ the topology of the Euclidean horizon  reduces to $L(2,1)\simeq S^3/\mathbb Z_2$, and, when also $p=0$, the above solutions provide the non-extremal index saddle carrying the contribution of the supersymmetric black lens of~\cite{Kunduri:2014kja,Kunduri:2016xbo} to the gravitational index. With $\big|n\big|>1$ and $p\in\mathbb Z$ subject to \eqref{eq:gcd3center}, instead, we are describing an infinite class of Euclidean supersymmetric black lenses with horizon $L\left(2\big|n\big|,1 - 2w_0^2\right) \simeq S^3/\mathbb Z_{2|n|}$. 
\end{itemize}

The on-shell action for these black ring and black lens candidate saddles follows from the analysis of~\cite{Cassani:2025iix,Colombo:2025yqy,Cassani:2026teb}. Assuming $p_1^2=1$, we have:
\begin{equation}
\label{eq:3centergeneral}
\begin{aligned}
\hatI &= \frac{\pi}{4G_5}\Biggl[C_{IJK} \frac{\check \varphi^I\,\check \varphi^J\,\check \varphi^K}{n\,\check \omega_1\,\check \omega_2} - C_{IJK}\frac{\left(p_1\check\varphi^I +\check\omega_2 \tilde q_1^I\right)\left(p_1\check\varphi^J +\check\omega_2 \tilde q_1^J\right)\left(p_1\check\varphi^K +\check\omega_2 \tilde q_1^K\right)}{n\,\check\omega_2\left(p_1 \,\check\omega_1 + \left(p_1 -1\right)\check\omega_2\right)}\Biggr] + 2\pi\ii \hatPsi 
\,,
\end{aligned}
\end{equation}
where 
\begin{equation}
\check\varphi^I = n\varphi^I -2\pi \ii \,\tilde q^I\,,\qquad \check\omega_1 = n\omega_1 + 2\pi\ii\left(1-n-p\right)\,,\qquad \check\omega_2 = n\omega_2 + 2\pi \ii p\,,
\end{equation}
and $\tilde q_1^I$ denotes the bubbling potential supported by the non-trivial topology outside the horizon.

As in \eqref{eq:2centregeneral}, a phase $\hatPsi$ independent of the chemical potentials, that renders the limit $n\to 0$ finite, has been introduced. We write it as 
\begin{equation}
    \hatPsi = \frac{\pi}{4nG_5}\Biggl[\left(1-p_1\right)w_0^2w_0^3 \,C_{IJK}\tilde q^I\tilde q^J\tilde q^K -  w_0^2 C_{IJK}\left(3\tilde q_1^I \tilde q^J \tilde q^K - 3p_1\,p\,\tilde q_1^I \tilde q_1^J \tilde q^K +p^2\tilde q_1^I \tilde q_1^J \tilde q_1^K \right)\Biggr]\,.
\end{equation}
This is controlled by the integers $w_0^2,w_0^3$ defined by \eqref{eq:w233center}. 
Although a rigorous derivation of this phase may be possible within the formalism of~\cite{Colombo:2025yqy}, we have not carried out a corresponding explicit computation. Its value is instead inferred from the requirement that the limit $n\to 0$ be smooth.

\paragraph{Horizonless soliton.}
As explained in section~\ref{sec:bubbling_saddles}, the limit $n\to 0$ gives a three-center horizonless soliton, where the Euclidean horizon is replaced by a spacelike \emph{bubble} with topology $S^1\times$ spindle. In this limit, also the bolt associated to the rod ${\cal I}_1$ reduces to $S^1\times$ spindle. This geometry, then, provides the Euclidean version of a class of horizonless solutions falling into the larger classification considered e.g.\ in~\cite{Bena:2005va,Bena:2007kg,Gibbons:2013tqa}. Taking into account the contribution from the phase controlled by the integers $w_0^{2,3}$, the general action \eqref{eq:3centergeneral} remains finite in this limit, giving
\begin{equation}
\label{eq:3centersoliton}
\begin{aligned}
    \hatI_{n\to 0} \,&=\, \frac{\pi}{4G_5}\Biggl[3C_{IJK}\left(\frac{\tilde q^I \tilde q^J}{1-p} + \frac{\left(p-p_1\right)p_1\left(p_1\tilde q^I - p\tilde q_1^I\right)\left(p_1\tilde q^J - p\tilde q_1^J\right)}{\left(1-pp_1\right)^2} \right)\frac{\varphi^K}{p}
    \\[1mm]
    \,&+\, C_{IJK} \Biggl(3\left(\tilde q^I \tilde q_1^J \tilde q_1^K- \tilde q^I \tilde q^J \tilde q_1^K\right) + p\left(p-2p_1\right) \tilde q_1^I \tilde q_1^J \tilde q_1^K +\frac{2p\left(p_1-1\right)}{\left(1-p\right)^2}\tilde q^I \tilde q^J\tilde q^K \Biggr) \frac{\omega_2}{\left(1-pp_1\right)^2}\Biggr]\,.
    \end{aligned}
\end{equation}

\paragraph{AdS$_3$ holography.} Uplifting to type IIB supergravity as in section~\ref{sec:STU} and taking the decoupling limit of section~\ref{sec:decoupling} yields three-center geometries with AdS$_3\times S^3$ asymptotics, satisfying the boundary conditions that are necessary for contributing to the elliptic genus \eqref{eq:ellipticgenus}, as explained in section~\ref{sec:AdS3xS3}.\footnote{A related zero-temperature decoupling limit was previously considered for Lorentzian BPS black rings in~\cite{Elvang:2004ds}, and for black lenses in~\cite{Kunduri:2016xbo}.} This also requires setting ourselves in a mixed ensemble where the charges $Q_1,Q_2$ are fixed. Both on-shell actions \eqref{eq:3centergeneral} and \eqref{eq:3centersoliton} remain finite in the decoupling limit. Their expression in the mixed ensemble is given by the transform \eqref{eq:changeensemble}. Applying it to~\eqref{eq:3centergeneral} yields
\begin{equation}
\label{eq:3centerAdS3}
\begin{aligned}
    I \,&=\, \frac{\pi^2}{2nG_6}\tilde q_1^1\tilde q_1^2 \check\varphi^3 \frac{\left(p_1\check\varphi^3 + \tilde q_1^3\check\omega_2\right)}{\left(1-p_1\right)\check\varphi^3 + \tilde q_1^3\check\omega_1} + \frac{2G_6Q_1Q_2}{n\pi^2}\frac{\check\omega_1\left(p_1\check\omega_1 - \left(1-p_1\right)\check\omega_2\right)}{\left(1-p_1\right)\check\varphi^3 + \tilde q_1^3\check\omega_1}
    \\[1mm]
    &\quad  - \frac{\tilde q_1^1 Q_1 + \tilde q_1^2 Q_2}{n p_1}\frac{\check\omega_1\left(p_1\check\varphi^3 + \tilde q_1^3\check\omega_2\right)}{\left(1-p_1\right)\check\varphi^3 + \tilde q_1^3\check\omega_1} + 2\pi \ii \Psi\,,
    \end{aligned}
\end{equation}
with
\be
\begin{aligned}
\Psi &=\frac1n\Biggl[ \tilde q^1 Q_1 + \tilde q^2 Q_2 + \frac{\pi^2}{2G_6}\Biggl(\left(1-p_1\right)w_0^2 w_0^3 \,\tilde q^1 \tilde q^2 \tilde q^3
\\[1mm]
&\quad -w_0^2\Bigl[ \tilde q_1^1 \tilde q^2 \tilde q^3 + \tilde q_1^2 \tilde q^1 \tilde q^3 +\tilde q_1^3 \tilde q^1 \tilde q^2 -pp_1 \left(\tilde q^1 \tilde q_1^2 \tilde q_1^3 + \tilde q^2 \tilde q_1^1 \tilde q_1^3 + \tilde q^3 \tilde q_1^1 \tilde q_1^2\right) + p^2\tilde q_1^1 \tilde q_1^2 \tilde q_1^3\Bigr]\Biggr)\Biggr]\,.
\end{aligned}
\ee
The transform of \eqref{eq:3centersoliton} instead gives
\begin{equation}
\begin{aligned}
    I \,&=\, \frac{\pi^2}{2G_6}\Biggl[\left(\frac{\tilde q^1 \tilde q^2}{1-p} + \frac{\left(p-p_1\right)p_1\left(p_1\tilde q^1 - p\tilde q_1^1\right)\left(p_1\tilde q^2 - p\tilde q_1^2\right)}{\left(1-pp_1\right)^2} \right)\frac{\varphi^3}{p}
    \\[1mm]
    \,&+\, \Biggl(\frac{|\epsilon_{IJK}|}{2}\left(\tilde q^I \tilde q_1^J \tilde q_1^K- \tilde q^I \tilde q^J \tilde q_1^K\right) + p\left(p-2p_1\right) \tilde q_1^1 \tilde q_1^2 \tilde q_1^3 +\frac{2p\left(p_1-1\right)}{\left(1-p\right)^2}\tilde q^1 \tilde q^2\tilde q^3 \Biggr) \frac{\omega_2}{\left(1-pp_1\right)^2}\Biggr]\,,
    \end{aligned}
\end{equation}
with 
\begin{equation}
    \begin{aligned}
        Q_1 &= - \frac{\pi^2}{2pG_6}\left[\frac{\tilde q^2 \tilde q^3}{1-p} + \frac{\left(p-p_1\right)p_1\left(p_1\tilde q^2- p\tilde q_1^2\right)\left(p_1\tilde q^3 - p\tilde q_1^3\right)}{\left(1-pp_1\right)^2} \right]\,,
        \\[1mm]
        Q_2 &= - \frac{\pi^2}{2pG_6}\left[\frac{\tilde q^1 \tilde q^3}{1-p} + \frac{\left(p-p_1\right)p_1\left(p_1\tilde q^1- p\tilde q_1^1\right)\left(p_1\tilde q^3 - p\tilde q_1^3\right)}{\left(1-pp_1\right)^2} \right]\,.
    \end{aligned}
\end{equation}
The latter expressions may then be solved e.g. for $\tilde q^1,\tilde q^2$ in terms of the fixed charges $Q_1,Q_2$.


\bibliography{decoupling.bib}
\bibliographystyle{JHEP}
      
\end{document}